\documentclass[letterpaper,twocolumn,10pt]{article}
\usepackage{titlesec}
\usepackage{siunitx}
\usepackage{url}
\usepackage{usenix}
\usepackage{epsfig}
\usepackage{booktabs}
\usepackage{xspace}
\usepackage{multirow}
\usepackage{listings}
\usepackage{amsmath}
\usepackage{subcaption}
\usepackage{caption}
\usepackage{filecontents}
\usepackage{xurl}
\usepackage{ulem}
\usepackage{wrapfig}
\usepackage{pifont}
\usepackage{comment}
\usepackage{xcolor}
\usepackage{threeparttable}
\usepackage{rotating}
\usepackage{xcolor}
\usepackage{graphicx}
\usepackage{colortbl}
\usepackage[misc]{ifsym}
\usepackage[linesnumbered, ruled]{algorithm2e}
\usepackage[affil-it]{authblk}
\usepackage[justification=centering]{caption}
\usepackage{balance}
\usepackage{pifont}
\usepackage{amssymb}

\makeatletter
\renewcommand\AB@affilsepx{\quad \protect\Affilfont}
\def\thanks#1{\protected@xdef\@thanks{\@thanks
        \protect\footnotetext{#1}}}
\makeatother

\definecolor{codegreen}{rgb}{0,0.6,0}
\definecolor{codegray}{rgb}{0.5,0.5,0.5}
\definecolor{codepurple}{rgb}{0.58,0,0.82}
\definecolor{backcolour}{rgb}{0.95,0.95,0.92}
\lstdefinestyle{mystyle}{
    backgroundcolor=\color{backcolour},   
    commentstyle=\color{codegreen},
    keywordstyle=\color{magenta},
    numberstyle=\tiny\color{codegray},
    stringstyle=\color{codepurple},
    basicstyle=\ttfamily\footnotesize,
    breakatwhitespace=false,         
    breaklines=true,                 
    captionpos=b,                    
    keepspaces=true,                 
    numbers=left,                    
    numbersep=5pt,                  
    showspaces=false,                
    showstringspaces=false,
    showtabs=false,                  
    tabsize=2
}
\SetKwFunction{FunctionName}{FunctionName}
\SetKwProg{Fn}{Function}{:}{}
\SetKwRepeat{Do}{do}{while}
\newcommand{\sysname}{P/D-Device\xspace}

\newcommand{\authorspace}{\hspace{-1ex + 1.5pt}}

\usepackage{tikz}
\usetikzlibrary{fadings}
\usetikzlibrary{shapes}
\usetikzlibrary{shadows.blur}

\begin{document}
\date{}

\title{\Large \bf \sysname: Disaggregated Large Language Model between Cloud and Devices}
\author[* \authorspace\thanks{$^*$Equally Contribution. $\dag$Work done during their internship at Huawei.}]{Yibo Jin}
\author[* \authorspace]{Yixu Xu}
\author[* \authorspace]{Yue Chen}
\author[* \authorspace]{Chengbin Wang}
\author[* \authorspace]{Tao Wang}
\author[ \authorspace]{Jiaqi Huang}
\author[ \authorspace]{Rongfei Zhang}
\author[ \authorspace]{Yiming Dong}
\author[$\dag$ \authorspace]{\\Yuting Yan}
\author[$\dag$ \authorspace]{Ke Cheng}
\author[$\dag$ \authorspace]{Yingjie Zhu}
\author[$\dag$ \authorspace]{Shulan Wang}
\author[ \authorspace]{Qianqian Tang}
\author[ \authorspace]{Shuaishuai Meng}
\author[ \authorspace]{Guanxin Cheng}
\author[ \authorspace]{Ze Wang}
\author[ \authorspace]{Shuyan Miao}
\author[ \authorspace]{Ketao Wang}
\author[ \authorspace]{Wen Liu}
\author[ \authorspace]{Yifan Yang}
\author[ \authorspace]{Tong Zhang}
\author[ \authorspace]{Anran Wang}
\author[ \authorspace]{Chengzhou Lu}
\author[ \authorspace]{Tiantian Dong}
\author[ \authorspace]{Yongsheng Zhang}
\author[ \authorspace]{Zhe Wang}
\author[ \authorspace]{Hefei Guo}
\author[ \authorspace]{Hongjie Liu}
\author[ \authorspace]{Wei Lu}
\author[ \authorspace]{Zhengyong Zhang}
\affil[ ]{Huawei Technologies Co., Ltd.}

\maketitle

\subsection*{Abstract}
Serving disaggregated large language models has been widely adopted in industrial practice for enhanced performance.
However, too many tokens generated in decoding phase, 
i.e., occupying the resources for a long time, 
essentially hamper the cloud from achieving a higher throughput.
Meanwhile, due to limited on-device resources, the time to first token (TTFT), 
i.e., the latency of prefill phase, increases dramatically with the growth on prompt length.
In order to concur with such a bottleneck on resources, 
i.e., long occupation in cloud and limited on-device computing capacity, 
we propose to separate large language model between cloud and devices.
That is, the cloud helps a portion of the content for each device, 
\textit{only in its prefill phase}.
Specifically‌, after receiving the first token from the cloud,
decoupling with its own prefill, the device responds to the user immediately for a lower TTFT.
Then, the following tokens from cloud are presented via a speed controller 
for smoothed TPOT (the time per output token),
until the device catches up with the progress.
On-device prefill is then amortized using received tokens
while the resource usage in cloud is controlled.
Moreover, during cloud prefill,
the prompt can be refined, using those intermediate data already generated, 
to further speed up on-device inference.
We implement such a scheme \sysname, and confirm its superiority over other alternatives.
We further propose an algorithm to decide the best settings.
Real-trace experiments show that
TTFT decreases at least 60\%, maximum TPOT is about tens of milliseconds,
and cloud throughput increases by up to 15x.

\section{Introduction}
\label{sec:introduction}

Serving large language models (LLMs~\cite{kaplan2020scaling,chatgpt4,deepseek2024v3,meta2025llama4}) 
in a disaggregated paradigm~\cite{patel2023splitwise,zhong2024distserve,jin2024pdserve,cunchen2024tetriinfer,gao2024attentionstore,qin2025fast,zuo2025serving}, 
has become a new trend,
where the prefill (P) and decoding (D) are deployed 
in different instances with disparate settings.
The prefill pursues lower time-to-first-token (TTFT) 
while decoding pursues larger batch size with tolerable time-per-output-token (TPOT).
Note that a lower TTFT results in quicker response to users 
while a larger batch size implies a higher throughput using the same resources.
In cluster-scale industrial practice,
serving disaggregated LLMs improves the system performance,
and various SLOs (service level objective~\cite{jin2024pdserve,qin2025fast}) are achieved for either P or D.

\begin{figure}[!t]
    \begin{subfigure}[h]{0.22\textwidth}
        \setlength{\abovecaptionskip}{2pt}
        \includegraphics[width=1.54in,height=1.265in]{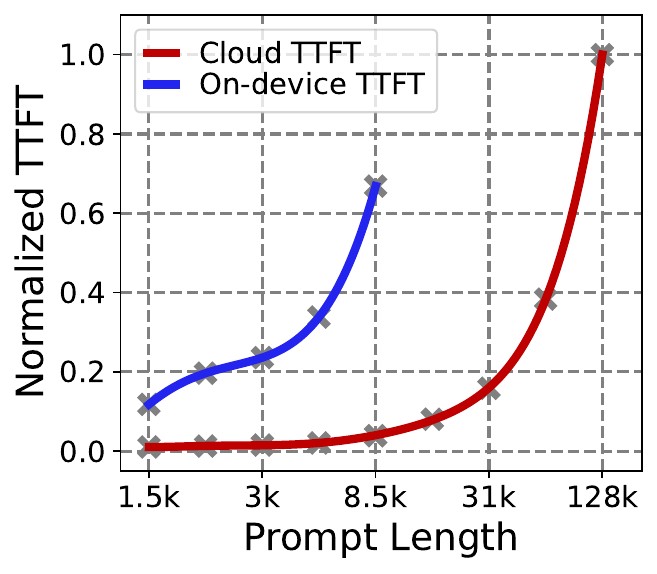}
        \caption{Growth on TTFT} 
        \label{fig:prefill_rel}
    \end{subfigure}
    \hfill
    \begin{subfigure}[h]{0.22\textwidth}
        \setlength{\abovecaptionskip}{2pt}
        \includegraphics[width=1.54in,height=1.265in]{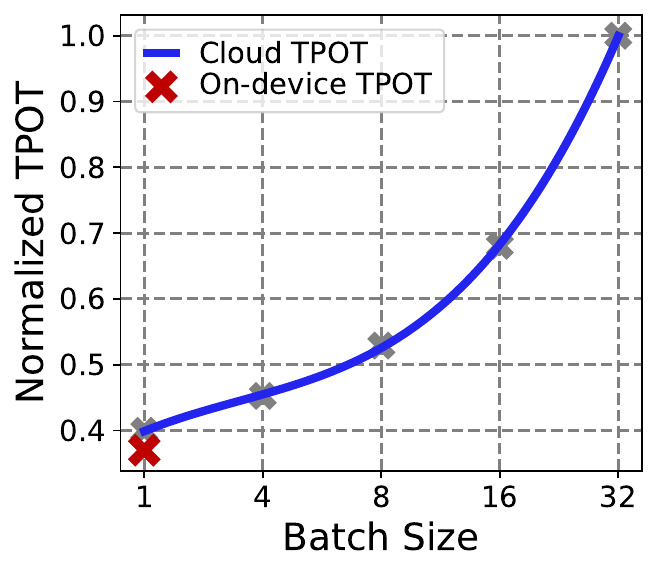}
        \caption{Growth on TPOT}
        \label{fig:decode_rel}
    \end{subfigure}
    \vspace{-7pt}
    \caption{TTFT and TPOT, in Cloud and on Devices}
    \label{fig:prefill_decode}
\end{figure}

\begin{table}[!t]
    \caption{Comparison on End-to-end Performance}
    \vspace{-8pt}
    \begin{tabular}{cccccc}
    \hline
    \multicolumn{1}{c|}{} & \multicolumn{1}{c|}{TPS} & \multicolumn{1}{c|}{TTFT} & \multicolumn{1}{c|}{TPOT} & \multicolumn{1}{c|}{Quality} & \begin{tabular}[c]{@{}c@{}}Desired\end{tabular} \\ \hline
    \multicolumn{1}{c|}{Device} & \multicolumn{1}{c|}{$\leq 1$} & \multicolumn{1}{c|}{>10s/8k} & \multicolumn{1}{c|}{\multirow{2}{*}{\shortstack{Tens\\ of ms}}} & \multicolumn{1}{c|}{Low} & TTFT$\downarrow$ \\ \cline{1-3} \cline{5-6} 
    \multicolumn{1}{c|}{Cloud} & \multicolumn{1}{c|}{$\propto 1/T$} & \multicolumn{1}{c|}{<1s/8k} & \multicolumn{1}{c|}{} & \multicolumn{1}{c|}{High} & $L \downarrow$ \\ \hline
    \multicolumn{6}{c}{$T=$ TTFT + TPOT$\cdot$($L$-1),\; $L$ Tokens in Total} \\ \hline
    \end{tabular}
    \vspace{-8pt}
    \label{compare_tab}
\end{table}

Recent studies also explore deploying on-device LLMs~\cite{XuQMHL18,YuanYCWYZLZMJWX24,xiao2025understanding}.
Unfortunately, either on-device LLMs or cluster-scale LLMs 
faces multiple challenges on resource usage.

First and foremost,
compared with the cloud inference, 
on-device TTFT growths dramatically with the increase on the prompt length (as illustrated in Fig.~\ref{fig:prefill_rel}), 
due to limited computing capacity~\cite{xiao2025understanding} 
(prefill is a compute-bound phase~\cite{zhong2024distserve,KamathPM0RP25}).
For those prompts with tens of thousands of tokens~\cite{google2025gemini25,openai2025gpt45,qwen3_2025} as the input,
on-device inference (models with several billion parameters~\cite{vaswani2023attentionneed,xue2024powerinfer2fastlargelanguage,yao2025efficient}) 
spends tens of seconds,
which causes a poor user experience (users have to wait long time for the first token).
Due to limited memory (CPU and NPU share the SoC memory~\cite{zhang2022benchmarking,xu25fastllm,kirinchip}),
the sizes of LLMs deployed are actually restricted.
Although some techniques like the sliding window attention~\cite{child19swa,beltagy2020longformer,xu2024ondevice,fu25swa} 
and quantification~\cite{park2024anyprecision,laskaridis2024meltingpoint,lin2024awqactivation,liu2024mobilellm} are adopted,
the device fails to maintain a resident instance for LLM serving.
Then, it is urgent to speed up on-device prefill (e.g., with the help of the cloud under the authorization).

Meanwhile,
long occupation per LLM request hampers the cloud from achieving a higher throughput.
The tokens are generated in an autoregressive manner~\cite{vaswani2023attentionneed},
where predicting the next token based on the previous ones.
Except for TTFT, TPOT is about tens of milliseconds~\cite{qin2025fast,zuo2025serving}.
With the growth on output tokens, the overall latency of decoding is considerable.
For example, 30ms per token multiplying 100 tokens leads to 3 seconds duration.
For summary task or document QA task with longer outputs, 
the timespan (just the decoding) may reach tens of seconds.
During such time period, related decoding instances are occupied
(those requests in the same batch share the instances).
Even equipped with tens of thousands of NPUs (or GPUs),
the throughput, i.e., transactions per second (TPS), is actually limited.
As shown in Table~\ref{compare_tab}, 
the throughput is inversely proportional to averaged latency. 
Ideally, the cloud prefers to serve requests with short outputs.

It is gratifying that
device and cloud are almost matched on TPOT
(device serves only one user while cloud serves a large batch), 
as in Fig.~\ref{fig:decode_rel}.
Note that decoding phase outputs one token per time, based on previous cache generated 
(i.e., KVCache~\cite{delcorro2023skipdecode,ge2024modeltell}).
It is promising to 
\textit{combine both faster prefill in cloud and almost matched decoding on devices},
to improve the overall system performance.

Unfortunately, it is challenging to exchange the intermediate data 
between cloud and devices, within short time interval.
Even if the model deployed in cloud is the same as that on devices,
the volume of the KVCache to be transferred is considerable (often GB sizes~\cite{strati2024dejavu,yang2024pyramidinfer}).
It is hard to ensure both time consumption on compression (preferred to be milliseconds)
and the data volume to be transferred (KB size is desirable).
Some works have studied speculative inference 
via frequent cloud edge collaboration~\cite{HaoJJ0C24,xie2025novelhatshaped,venkatesha2025fastcost},
where the data are regularly sent by devices and verified in the cloud.
All the candidates have to be verified timely
to ensure the averaged TPOT is smaller than that on devices.
Then, about tens of milliseconds are left for the round trip.
Once the network jitters occur~\cite{DahmouniGS12,MesbahiD16}, the averaged TPOT of verified tokens may not be controlled.
Therefore, the collaboration mechanism should be well orchestrated
under tolerable communication load (for realistic usage, considering both frequency and volume).

Existing research falls insufficient for these challenges.
Some works studied intra-cluster serving system~\cite{agrawal2023sarathi,patel2023splitwise,zhong2024distserve,jin2024pdserve,gao2024attentionstore,cunchen2024tetriinfer,ye2024chunkattention,ZengGLYSHWZ00Q24,brakel2024modelparallelism,deepseek2024v3,qin2025fast,zuo2025serving,0002HQZYCZH025,qiao2025swiftkv,wang2025flexsp,lin2025apex}.
Others focused on on-device LLM~\cite{zhang2022benchmarking,park2024anyprecision,laskaridis2024meltingpoint,liu2024mobilellm,xiao2025understanding,xu25fastllm}.
And the rest investigated the collaboration~\cite{LiZZC20,LaskaridisVALL20,ZhangZQWJL21,AlmeidaLVLL22,ChenLZPWHZ24,HaoJJ0C24,zhang2024edgeshard,yao2024gkt,venkatesha2025fastcost}.
However, few of them has considered disaggregated LLM between cloud and devices,
with cloud TTFT, KB-size transfer (for long context), smoothed TPOT, improved quality (compare with on-device one) and increased throughput.

In this paper, we propose a new collaboration scheme \sysname,
in which the cloud helps a portion of the content for each device, \textit{only in its prefill phase}.
Via assisting a portion of the content, 
the cloud controls the resources usage per request, to avoid long occupation on decoding.
As a result, the cloud is more likely to achieve a higher throughput.
At the same time, the tokens received per device are well designed for multiple purposes.
With token-level help from cloud, 
each device has the chance to respond the users quickly (i.e., using those tokens already received), 
as though they were generated by the device itself.
Via decoupling the display from the inference,
long on-device prefill can be essentially amortized,
in which the TPOT among assisted tokens is moderately enlarged.
Note that the reading speed of a human is about hundreds of words per minute~\cite{BRYSBAERT2019104047} 
(i.e., TPOT within hundreds of milliseconds).
During the amortization,
the device can rapidly catch up with the progress (generating the same number of tokens received), 
since on-device decoding and cloud decoding are almost matched.
Once the device gets rid of the prefill, it is capable of inferring the following tokens itself.
Further, if the model deployed in cloud is larger (with the same distribution),
the tokens assisted can be regarded as the ``ground truth'',
and be used to correct on-device decoding 
(if the difference occurs between tokens at the same position).

We further explore to refine the prompt for on-device prefill (for a shorter prompt).
During cloud prefill, the intermediate data, generated per attention layer, 
actually implies the relative importance among the input tokens (e.g.,~\cite{Zhang00CZC0TRBW23,LiHYVLYCLC24}).
We use it to filter the content, based on a desired ratio.
Note that all the information has already been generated during cloud prefill.
Then, with the first token transferred from cloud to the device,
refined prompt can be also delivered (via the mask in KB sizes).
Upon received shorter prompt,
the device triggers the prefill, leading to a faster TTFT.
Essentially, via refined prompt,
the model specifications between cloud and devices are decoupled.
\textit{It is possible to deploy a more powerful model in cloud to help the devices.}
With the growth on prompt length (long context),
it is more necessary to select the most important content for devices, to balance both efficiency and quality.
Moreover, to ensure the semantic coherence, the refinement is performed via sentence-level selection.

We implement \sysname in our prototype with real LLMs deployed in cloud and on devices.
To further balance TTFT, TPOT and the inference quality,
we formulate the procedure of \sysname
and propose an algorithm to decide the best settings 
(i.e., the refinement ratio for prompts and the number of tokens assisted by cloud).
Via real-trace experiments,
the superiority of \sysname over other alternatives is confirmed,
with cloud-level TTFT (user-perceived one decreases at least 60\%),
smoothed TPOT (maximum is tens of milliseconds, no harm to reading),
and higher throughput (increases by up to 15x, compared with cloud inference).
As for the quality, even deployed the same model in cloud and on devices,
the score under the collaboration also improves due to precise selection
(a large model can be used for further improvement).

\vspace{-1pt}
\section{Background and Motivation}
\vspace{-3pt}
\label{background}
This section analyzes existing serving systems and illustrates their bottlenecks on resource usage (both cloud and devices).
\clearpage

\subsection{LLM Serving System}
The cloud contains a vast nodes, in which part of the nodes (equipped with several NPUs (or GPUs) per node) 
are responsible for providing various LLM services, 
to form existing serving system (including the network, load balancer, disaster recovery components, etc.).
In contrast, due to limited resource (e.g., the memory), 
the device (smart phones, tablets, etc.) fails to maintain a resident LLM instance.
Instead, the inference is triggered in an on-demand manner.

\subsubsection{Cluster-scale Serving}

\textbf{Disaggregation:}
Traditionally, each node serves one or more LLM instances (via docker container) with fully functionality of prefill and decoding.
With the growth on the model sizes~\cite{abs-2303-10845,deepseek2024v3,meta2025llama4,qwen3_2025} 
and separate SLOs~\cite{qin2025fast} (shorter TTFT in prefill and larger batch in decoding),
single node is no longer sufficient.
Instead, the LLM instance is supported by multiple nodes (e.g., one container per node),
where each container is responsible for a part of inference.
For example, the disaggregation of prefill and decoding has become a new trend,
where the prefill and decoding phases are deployed in different containers~\cite{jin2024pdserve} with disparate settings.
Since the decoding requires the intermediate data 
(i.e., KVCache~\cite{delcorro2023skipdecode,ge2024modeltell}) 
generated in prefill for follow-up tokens,
KVCache is then transferred over RoCE (NPU-to-NPU directly, RDMA over Converged Ethernet).
Even in the same phase,
multiple nodes are also adopted (e.g., one node for part of MoE computations).

\textbf{Inefficient Decoding:}
Via disaggregated LLM, the serving system pursues lower TTFT and larger decoding batch.
Although the TPOT is only tens of milliseconds~\cite{qin2025fast,zuo2025serving},
the duration of decoding may last for a long time.
The overall latency of decoding is TPOT multiplying the number of tokens generated
(hundreds of tokens generated leading to tens of seconds).
For the scenarios with plenty of output tokens (e.g., summary and document QA~\cite{meta2025llama4}),
the decoding nodes are continuously occupied~\cite{continuousbatching}.
As a result, those occupied decoding nodes fail to serve further requests,
until one request in the batch completes its inference (e.g., generates end-of-token label).
Actually, continuously occupied nodes restrict the throughput, even equipped with tens of thousands of NPUs
(i.e., the number of requests treated simultaneously is limited).
Ideally, the cloud prefers to treat each request in an efficient manner 
(i.e., fewer tokens per request for higher throughput).

\subsubsection{On-device Deployment}

\textbf{On-demand Trigger:}
On-device inference mainly relies on the system-on-a-chip architecture (SoC)~\cite{kirinchip}, 
in which the integrated circuit combines most or all key components onto a single microchip.
Typically, the SoC includes both CPU and GPU for central processing and graphics processing, respectively.
Moreover, equipped with NPU on SoC, the machine learning models can be further accelerated.
Unfortunately, all these processing units share the memory~\cite{zhang2022benchmarking,xu25fastllm} (typically several gigabytes).
It is unrealistic to keep one LLM instance maintained (or its tiny version) in the memory (also consider the power consumption).
Instead, only when the target application (APP) is triggered,
the LLM is loaded (scenario switch is implemented via LoRA~\cite{applelora,googlelora}).
Due to on-demand manner, the user or APP has to be waiting, 
until all the preparations are completed and then the inference starts.

\textbf{Long Prefill:}
As in Fig.~\ref{fig:prefill_decode}, 
the device and cloud are almost matched on decoding speed (about tens of milliseconds).
Note that one device only serves one user or several APPs per time 
while the cloud serves a large batch simultaneously.
Although the decoding phase outputs one token per iteration (autoregressive), 
it requires the KVCache generated by the device itself in prefill.
Due to limited resources (peak performance of NPU and the SoC memory), 
on-device TTFT is much longer than cloud TTFT.
We should mention here that, small LLMs (SLM) with several billion parameters are suitable for devices
(the cloud serves the model with up to hundreds of billions of parameters).
Meanwhile, those lightweight techniques are adopted for 
reduced computation and memory consumption on devices (sliding window, quantification, etc.).

\begin{figure}[!t]
    \centering
    \includegraphics[width=3.33in,height=1.61in]{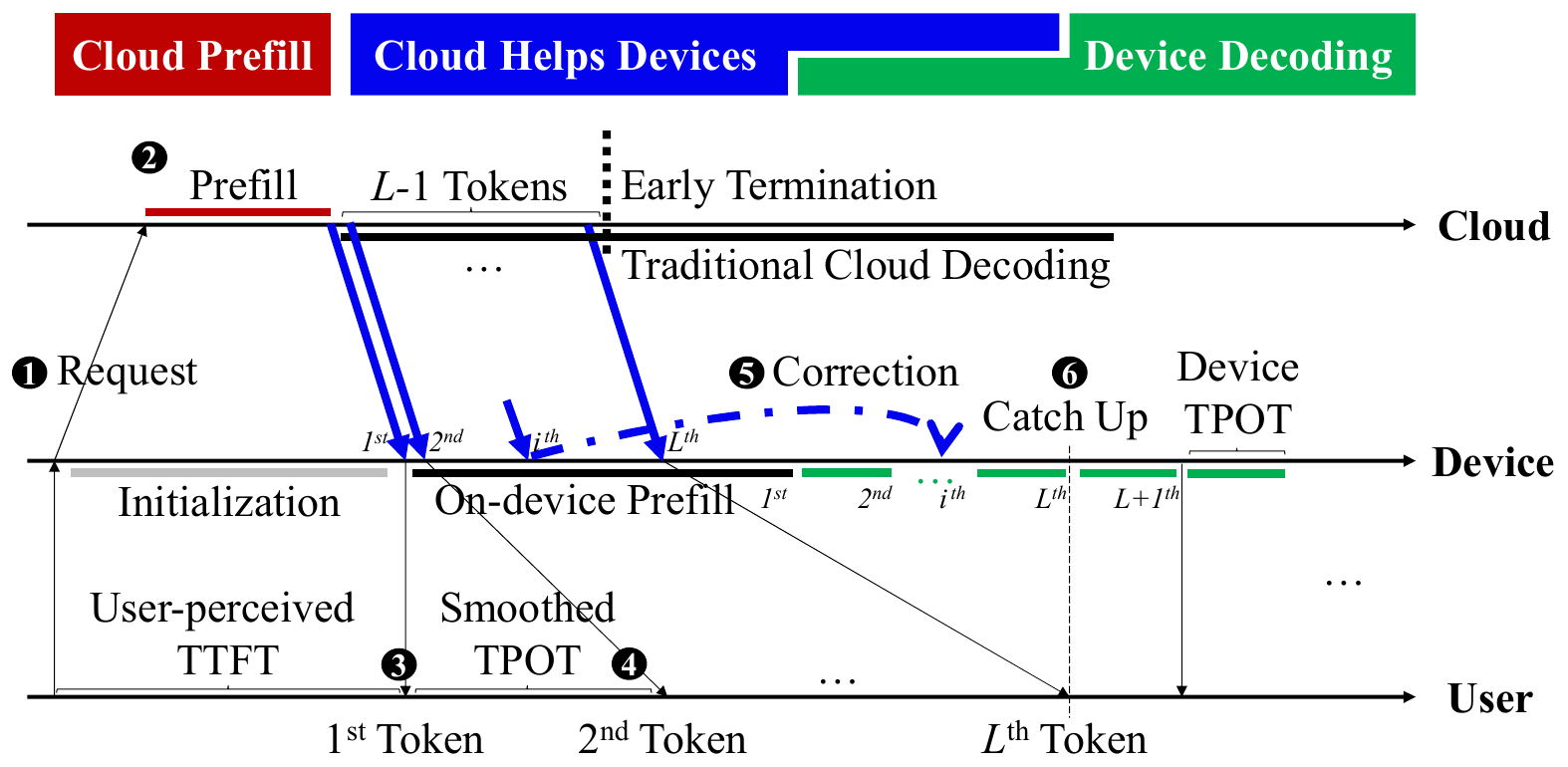}
    \vspace{-19pt}
    \caption{Overview of Cloud-assisted Inference}
    \label{fig:case_one}
    \vspace{-11pt}
\end{figure}

\subsection{Challenges and Opportunities}
The cloud prefers to serve LLM requests with less outputs (avoid long occupation on decoding) 
while the device prefers to infer with shorter on-device TTFT (avoid heavy computation in prefill).
It is promising to combine both faster prefill in cloud 
and almost matched decoding on devices (i.e., cloud helps devices),
to improve the overall performance.

However, it is challenging to exchange such large volume of intermediate data 
between cloud and devices within short time interval.
Although some recent works have studied cloud-device collaborations to improve the performance,
it is hard to achieve ensured efficiency, especially for long context.

\subsubsection{Requirements on Efficiency for Long Context}

\textbf{Real-time Collaboration:}
Cloud TTFT is about hundreds of milliseconds.
Ideally, the exchange of the intermediate data is desired to be completed within tens or hundreds of milliseconds,
to facilitate the collaboration.
Unfortunately, the network jitters are unavoidable during the round trip~\cite{DahmouniGS12,MesbahiD16}.
Further, the volume of KVCache generated in prefill is about GB sizes~\cite{strati2024dejavu,yang2024pyramidinfer}.
It is unrealistic to ensure real-time exchange for such volume of data.
Instead, some works have studied the token-level collaboration.
Note that on-device inference is often accelerated 
via speculative decoding~\cite{HaoJJ0C24,xie2025novelhatshaped,venkatesha2025fastcost}.
Several tokens are exchanged per time and the cloud performs related verification.
To ensure the averaged TPOT of verified tokens is smaller than that on devices,
such collaboration requires a tighter deadline.
For example, if the LLM generates 5 tokens with the TPOT of 35 milliseconds (ms),
i.e., the SLO: 35ms per token.
Via speculative decoding (e.g., acceptance rate is 60\% per 5 tokens), to ensure the same SLO per token,
5 * 60\% = 3 tokens have to be verified within 3 * 35 = 105ms.
If on-device SLM generates one token using 8ms and the cloud verifies 5 tokens using 30ms,
only 105 - 8 * 5 - 30 = 35ms is left for RTT (hard to be achieved).
Thus, even for token-level collaboration, the mechanism should be well studied.

\textbf{Scalability on Long Context:}
Each communication has the potential to break the tight deadline, which is unfriendly to the collaboration.
With the growth on the outputs, frequent communications in speculative decoding further increase the uncertainty
(and the cloud has to maintain the connection with related KVCache kept).
Instead, it is preferred to fully utilize the already existing round trip 
(e.g., the cloud only responds once and transfers sufficient information).
Some works like GKT~\cite{yao2024gkt} enable the knowledge distillation in cloud as the guidance,
in which such information is appended to the prompt as the clue.
However, for long context as the input, the device still suffers from unacceptable prefill.
Therefore, the collaboration mechanism should also control the prompt length for devices
(on-device prefill is a must to generate KVCache for follow-up tokens).
In comparison, the texts (prompt, tokens, etc.) are more suitable than raw KVCache,
since the KVCache grows much faster, proportional to the prompt length.
Essentially, both input (long on-device prefill) and output (long cloud occupation) should be controlled.

\subsubsection{Token-level Assist during On-device Prefill}
\textbf{Opportunity:} As in Fig.~\ref{fig:case_one},
we propose to combine faster prefill in cloud and almost matched decoding on devices.
During the device's prefill, the cloud helps a portion of the content 
(i.e., refined prompt and controlled number of tokens).

\textbf{Cloud Prefill:}
Compared with long on-device TTFT (several seconds or more),
cloud TTFT spends only hundreds of milliseconds.
Thus, after removing the sensitive information 
(due to privacy protection, phone number, real name, etc.),
the device sends the prompt 
(raw text with formatted settings and questions, as a request, step \ding{182}) to the cloud.
The cloud (i.e., LLM serving instances) performs its prefill (step \ding{183}) 
and responds the first (1$^{st}$) token back to corresponding device.
To pursue quick response to users,
the 1$^{st}$ token is presented immediately (i.e., user-perceived TTFT  equals to cloud TTFT plus RTT, step \ding{184}).
We should mention here that, the technique like private cloud compute (PCC~\cite{applelora})
proposed ensures the private AI processing in cloud (under user permission).

\textbf{Cloud Helps Devices:}
Instead of completely cloud decoding or on-device decoding,
after the 1$^{st}$ token,
the cloud continuously generates decoding tokens.
To avoid extra communications (fully utilize the SSE feedback),
the maximum number of decoding tokens is pre-defined (studied later), 
which actually controls the early termination in cloud (avoid long occupation).
With the follow-up tokens generated and transferred,
the device appropriately slows down their display,
as if they were generated by device itself 
(with smoothed TPOT, step \ding{185}, e.g., tens of milliseconds, to amortize long on-device prefill).
If the models deployed in cloud and on devices are the same,
the tokens assisted can be adopted upon the policy.
Otherwise, the tokens from a larger model in cloud (the same distribution required) are actually the ``ground truth''.
Once the token (produced by device) is different from the one received (at the same position),
cloud one is presented and used as the input (step \ding{186}) for next token generation 
(optional, like the speculative decoding, use the ones from cloud).

\textbf{Device Decoding:}
After the device catching up the tokens received (step \ding{187}),
follow-up tokens are generated by itself.

\begin{figure}[!t]
    
    \begin{subfigure}[h]{0.22\textwidth}
        \setlength{\abovecaptionskip}{-1pt}
        \includegraphics[width=1.53in,height=1.265in]{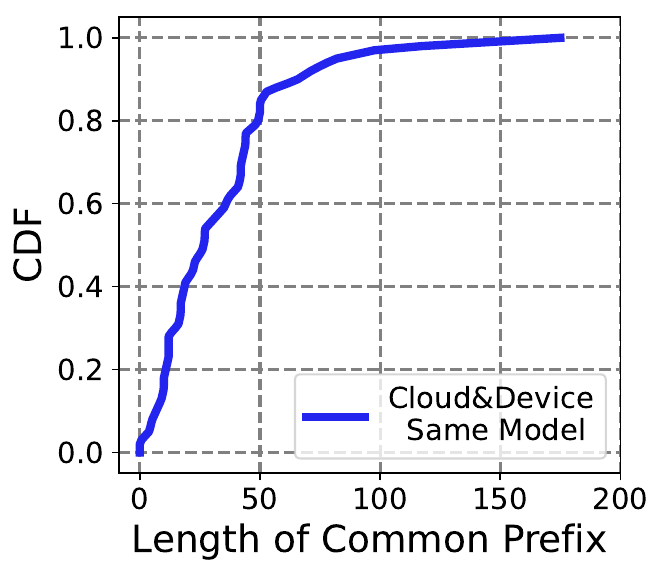}
        \caption{Common Prefix between\\Cloud and Device Outputs}
        \label{fig:case1_1}
    \end{subfigure}
    \hfill
    \begin{subfigure}[h]{0.22\textwidth}
        \setlength{\abovecaptionskip}{-1pt}
        \includegraphics[width=1.53in,height=1.265in]{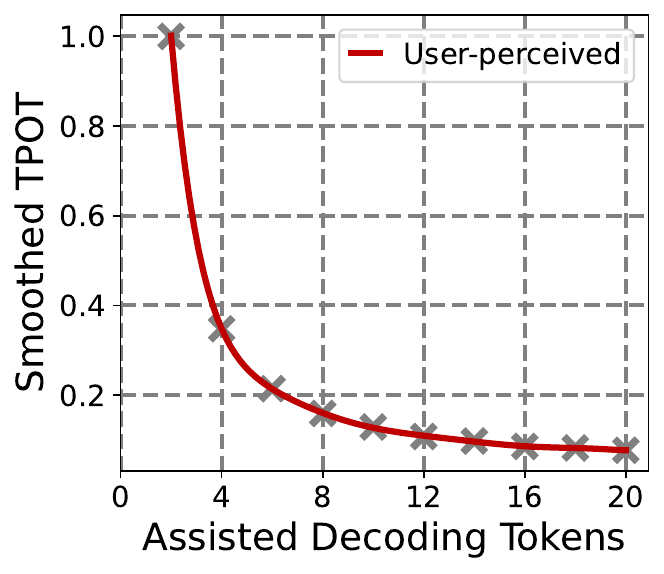}
        \caption{Smoothed TPOT under Various Assisted Tokens}
        \label{fig:case1_2}
    \end{subfigure}
    \vspace{7pt}

    \begin{subfigure}[h]{0.22\textwidth}
        \setlength{\abovecaptionskip}{-1pt}
        \includegraphics[width=1.53in,height=1.265in]{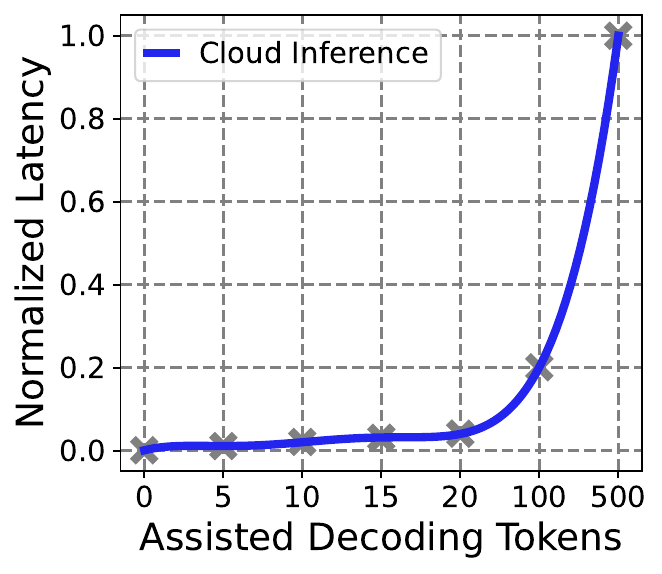}
        \caption{Cloud Latency under Various Assisted Tokens} 
        \label{fig:case1_3}
    \end{subfigure}
    \hfill
    \begin{subfigure}[h]{0.22\textwidth}
        \setlength{\abovecaptionskip}{-1pt}
        \includegraphics[width=1.53in,height=1.265in]{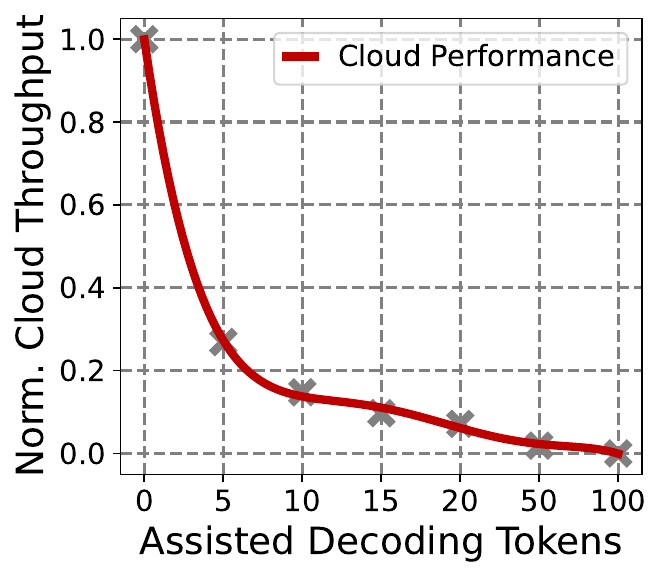}
        \caption{Cloud Throughput under Various Assisted Tokens}
        \label{fig:case1_4}
    \end{subfigure}

    \vspace{-5pt}
    \caption{Results under Various Assisted Tokens}
    \vspace{-9pt}
\end{figure}

\textbf{Analysis on Assistance:}
Assisted tokens can be used to amortize long on-device prefill 
(TTFT includes prefill; traditionally, on-device TTFT$\approx$prefill$_{d}$).
It is about seconds or even tens of seconds.
Although TPOT is tens of milliseconds,
on-device prefill is unacceptable.
The amortization is actually the following format
($L$ tokens assisted, i.e., except for the 1$^{st}$ token, 
$L-1$ decoding ones to amortize prefill$_{d}$):
\begin{equation}
	\text{prefill}_{d}\text{ + \small TPOT}_{d}*\text{(}L\text{-1)} = \text{\small TTFT}_{c}\text{ + \small TPOT}_{smooth}*\text{(}L\text{-1)},\label{amortized}
\end{equation}
where notation ``d'' refers to on-device inference
and ``c'' refers to cloud inference.
TTFT$_{c}$ (i.e., cloud TTFT) is about hundreds of milliseconds.
As in Fig.~\ref{fig:case1_1},
with the growth on $L-1$ (the decoding tokens assisted),
TPOT$_{smooth}$ (i.e., smoothed TPOT) drops dramatically.
Using 20 decoding tokens assisted,
smoothed TPOT is about one tenth of that using 2 tokens,
and is more closer to the value of TPOT$_{d}$ (i.e., on-device TPOT).
Since TTFT$_{c}$ is far smaller than prefill$_{d}$,
TPOT$_{smooth}$ is always larger than TPOT$_{d}$.
Note that appropriately enlarging user-perceived TPOT is acceptable
(i.e., match the reading speed of a human, about hundreds of words per minute~\cite{BRYSBAERT2019104047}).

Fig.~\ref{fig:case1_1} shows the common prefix between cloud and device outptus 
(cloud uses the original prompt while device uses the refined one (studied later)),
in which the models deployed are the same.
Although some lightweight techniques are adopt on devices,
the average length of the common prefix is about tens of tokens.
Such common prefix implies that
the tokens assisted can be used in advance (display and amortization), prior to the device decoding.
As in Fig.~\ref{fig:case1_2},
via 20 tokens assisted, smoothed TPOT is about tens of milliseconds,
in which the device prefill has been amortized.
Even if the tokens are different at the same position,
the device has the chance to conduct the correction:
correct its own inference (use cloud one for next generation)
or correct the display (use the one generated by itself).
If the models used in cloud are larger than that on devices (the same distribution required),
assisted tokens can be regarded as the ``ground truth''.
Note that these two bad cases frequently occur on devices:
repeated generation (produce repeated tokens) 
and semantic inconsistency (the outputs are quite different from the user intent).

Fig.~\ref{fig:case1_3} shows the latency 
under various decoding tokens assisted by the cloud.
Compared with all decoding tokens generated in cloud (e.g., 500 tokens),
the early termination just allows 20 decoding tokens assisted.
And the overall decoding is about 1 second
(instead, 500 tokens require tens of seconds duration).
Via offloading the heavy decoding phase to devices 
(i.e., cloud only generates tens of tokens instead of hundreds of tokens or more),
the throughput improves dramatically, as shown in Fig.~\ref{fig:case1_4}.
The throughput is about inversely proportional to averaged latency.
Reducing the averaged latency of LLM requests in cloud actually improves the throughput
(cloud serves plenty of scenarios and models).

\subsubsection{Prompt-level Assist after Cloud Prefill}
\textbf{Opportunity:} Token-level assist uses decoding tokens from cloud 
to amortize long on-device prefill.
It is further promising to speed up on-device prefill, 
using the intermediate data already generated during the cloud prefill.
Note that faster on-device prefill leads to earlier start of decoding.
Along with the feedback of 1$^{st}$ token,
the cloud can further help the device prefill with more details (e.g., shorter prompt).

\textbf{Cloud Refines Prompts:}
On one hand, for those scenarios 
(e.g., summary, document QA, etc.) with plenty of tokens,
cloud prefers to serve the requests with less outputs.
On the other hand, those scenarios often require long context as the inputs,
including both format part (background, settings, etc.) 
and main content part (articles, blogs, comments, etc.),
whose contents can be inherently refined.
For example, after summarizing each paragraph,
the summary of an article can be obtained by combining all these refined information.

The transformer paradigm, actually the attention mechanism,
essentially determines the importance of each item (i.e., the token) in a sequence.
Specifically, for standard scaled dot-product attention~\cite{vaswani2023attentionneed},
the inputs are represented into three tensors (\textbf{Q}, \textbf{K}, \textbf{V}),
where \textbf{Q} represents what you are looking for,
\textbf{K} represents the reference of relevant information,
and \textbf{V} represents the actual semantic meaning.
Such attention is
\begin{equation}
	Attention(\text{\textbf{Q}}, \text{\textbf{K}}, \text{\textbf{V}}) = softmax({\text{\textbf{Q}}\text{\textbf{K}}^{\intercal}}/{\sqrt{h}})\text{\textbf{V}},\nonumber
\end{equation}
where $h$ is the hidden size.
It calculates the closeness between the query and each key-value pair.
The dot product of \textbf{Q} and \textbf{K} indicates the similarity
(a.k.a. attention score).
Intuitively, if the similarity is higher for a key,
its corresponding value is more likely to be chosen (among all tokens).
For the scope of input tokens (all tokens naturally contain the input ones),
their score values indicate relative importance,
which can be used as the guidance to refine the content (for shorter prompt).

\begin{figure}[!t]
    \begin{subfigure}[h]{0.22\textwidth}
        \setlength{\abovecaptionskip}{-1pt}
        \includegraphics[width=1.53in,height=1.265in]{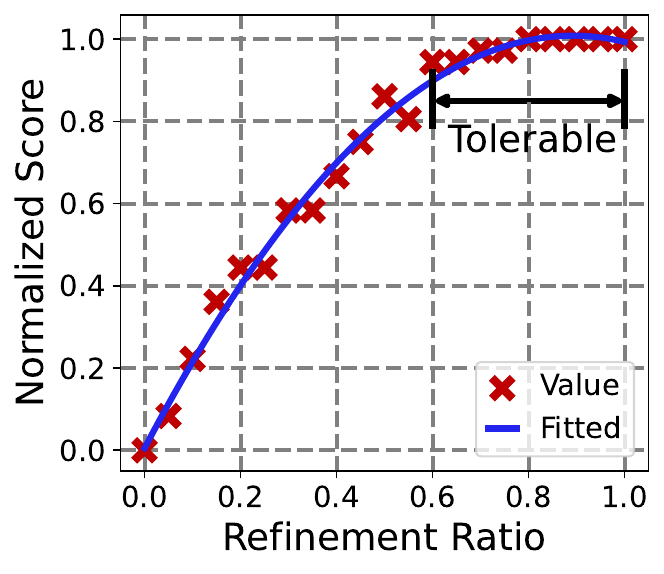}
        \caption{Inference Quality under Various Refinement Ratio}
        \label{fig:case2_1}
    \end{subfigure}
    \hfill
    \begin{subfigure}[h]{0.22\textwidth}
        \setlength{\abovecaptionskip}{-1pt}
        \includegraphics[width=1.53in,height=1.265in]{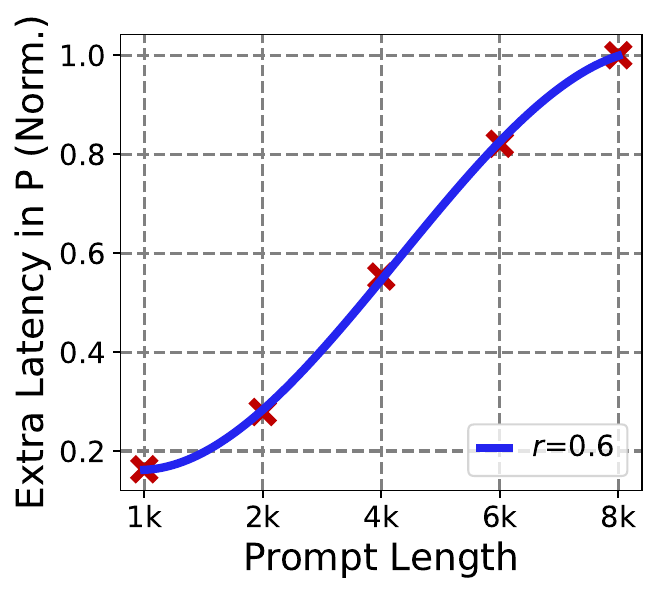}
        \caption{Extra Latency\\(Refinement and Compression)}
        \label{fig:case2_2}
    \end{subfigure}
    \vspace{-7pt}
    \caption{Results using Refined Prompts}
    \vspace{-8pt}
\end{figure}

\textbf{Analysis on Assistance:}
We use $r$ to represent the refinement ratio (ranging from 0 to 1),
which is calculated by the length of refined prompt dividing the length of original one.
With the decrease of the ratio, the score naturally drops.
However, there exists a tolerable range.
As in Fig.~\ref{fig:case2_1},
the quality drops at most several percent under 40\% decrease on prompt length (i.e., refinement ratio is 0.6).
Here, the model tested in cloud is the same as that on devices.
Further shown later, via precise selection, 
even using the refined prompt, the quality under collaboration is more likely to be improved
(most context contains redundant information).
Fig.~\ref{fig:case2_2} shows the extra latency involved
(at most several hundred milliseconds for 8k prompt).
With the decrease on prefill,
TPOT$_{smooth}$ decreases, upon Eq.~\ref{amortized}
(values remain unchanged on both two sides).

\textbf{Remarks:} Prompt-level assist is essentially orthogonal to token-level assist.
Without refined prompts,
token-level assist itself already enables 
the amortization for long on-device prefill.
No matter the token-level assist ($L \geq$ 1) is enabled or not,
the refined prompt can be simply applied
(i.e., piggybacking with the feedback of the 1$^{st}$ token).

\section{System Design and Implementation}

\subsection{System Overview}
\begin{figure}[!t]
    \centering
    \includegraphics[width=3.34in,height=1.65in]{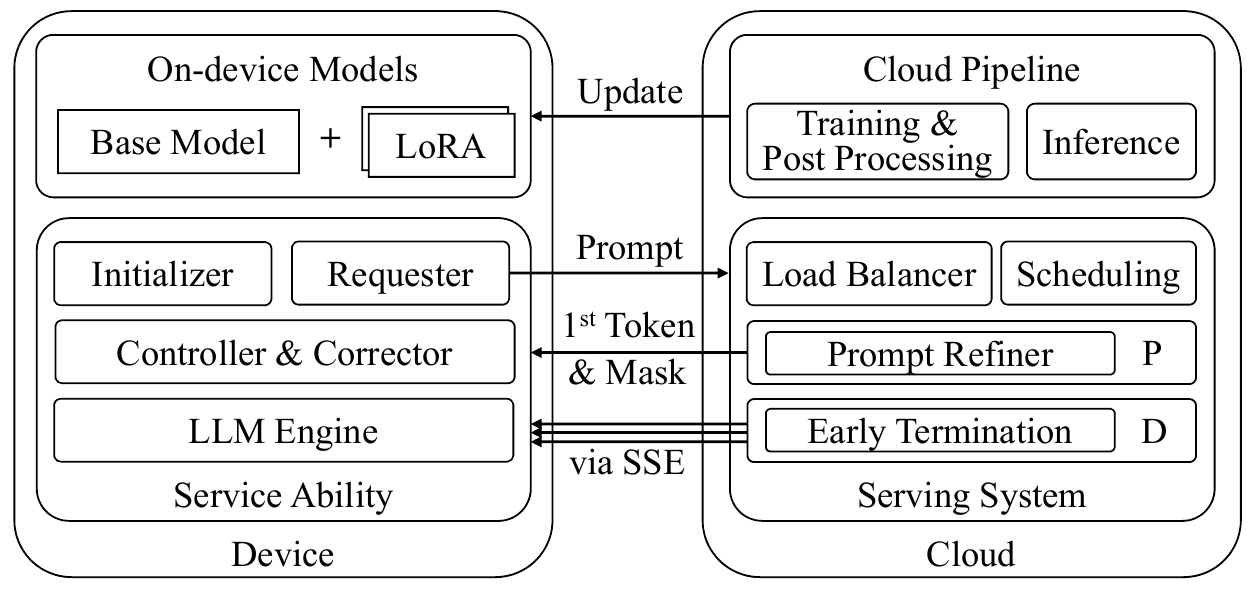}
    \vspace{-17pt}
    \caption{System Overview: Cloud-device Disaggregation}
    \label{fig:architecture}
    \vspace{-7pt}
\end{figure}

Fig.~\ref{fig:architecture} shows the system overview of \sysname,
in which the inference requires real-time collaboration between cloud and devices.
Meanwhile, the models are updated offline.

\sysname contains three main components as follows:

\textbf{Cloud Pipeline and Serving System:}
Due to limited resources on devices,
LLM is enabled via the combination of base model and LoRA,
in which both base model and LoRA are updated using cloud pipeline.
The pipeline covers both training and inference,
and further the post processing part, 
including conversion, pre-compilation, etc., for faster on-device inference.
The cloud manages the version matching information and 
dispatches corresponding version of base model and LoRA for various devices.
Note that the switching of diverse scenarios can be implemented via the switching of LoRA.

As illustrated in section \ref{sec:design_cloud}, the serving system is the core of cloud inference,
in which the load balancer receives all the requests and delivers them
to related services and instances.
All the instances are isolated via containers (elastically scaled out),
in which each container is responsible for part of the inference (prefill (P), decoding (D) or both of them).
Except for intra-cloud requests (triggered by the services already deployed in cloud),
the instances identify the ones from devices and 
enables the collaboration via the prompt refiner and early termination.
Here, the optimizer decides the best settings of prompt refiner and early termination, 
as illustrated in section \ref{sec:design_algorithm}.
Specifically, prompt refiner compresses the prompt,
using the intermediate data, already generated during prefill, 
and transfers a mask along with the feedback of the 1$^{st}$ token to devices.
The early termination completes cloud decoding to avoid long resource occupation.

\textbf{On-device Models and Service Ability:}
All the models received are stored on devices,
in which each pair of base model and LoRA is related to one specific scenario 
(summary for APP \textit{A}, continued writing for APP \textit{B}, etc.),
based on the configurations automatically generated during cloud pipeline.
Due to limited memory and battery,
maintaining the entire LLM for serving is unrealistic.
Instead, the service ability of on-device operating system triggers the initialization in an on-demand manner,
including prompt desensitization, weight loading, etc.
Here, the service ability is used to run tasks in the background, authorized by users or APPs.
After sending the request and receiving the feedback from cloud,
the service ability uses 
both token-level and prompt-level assist to amortize and accelerate on-device inference.

As illustrated in section \ref{sec:design_device},
the LLM engine executes the inference,
in which the lightweight techniques (attention implementation, quantification, etc.) are adopted,
due to limited computing capacity and memory.
The speed controller adjusts the TPOT to amortize on-device prefill 
while the corrector rectifies wrong tokens, using the ``ground truth'' from cloud.
Although the refined prompt postpones on-device prefill 
(i.e., till the moment the 1$^{st}$ token received),
due to shorter prompt length,
on-device prefill is actually decreased.

\textbf{Scheduling on Cloud-device Disaggregation:}
As shown in section \ref{sec:design_algorithm},
the optimizer in cloud determines the best settings for cloud-device collaboration (per scenario),
in which the refinement ratio and the number of decoding tokens assisted are included.
The decision keeps the balance between adequate cloud assist and controlled resource usage,
leading to minimized TTFT and smoothed TPOT (user-perceived).

\subsection{Cloud Assistance: Prompts and Tokens}
\label{sec:design_cloud}

\textit{Preliminaries:}
There are two execution modes, supported by Ascend~\cite{ascend}: static graph and dynamic graph.
Static graph mode performs pre-compilation for better performance
while the program is executed line by line (e.g., per each operator) in dynamic graph mode.
The former pursues extreme performance while the latter improves the development efficiency.
Without pre-compilation in static graph mode,
the inference has to compile and optimize the graph in an on-demand manner 
(may involve several minutes or more).
Via reusing the output of such compilation, 
the model initialization (for scale-out, scenario switching, etc.) is actually accelerated.
Furthermore, fused kernels are enabled for improved performance (attention, flash attention, etc.),
in which similar operations are combined and multiple streams are used.

During the execution,
all necessary intermediate data is stored in the HBM of GPUs or NPUs (e.g., KVCache).
Instead of moving the data per line or per operator to host memory for synchronization,
all the tensors we needed are obtained upon static graph mode,
in which we extend attention implementation and fetch them efficiently.
Note that all extra operations are executed in-place, 
instead of involving extra data copies.

\begin{figure}[!t]
    \centering
    \includegraphics[width=3.33in,height=1.24in]{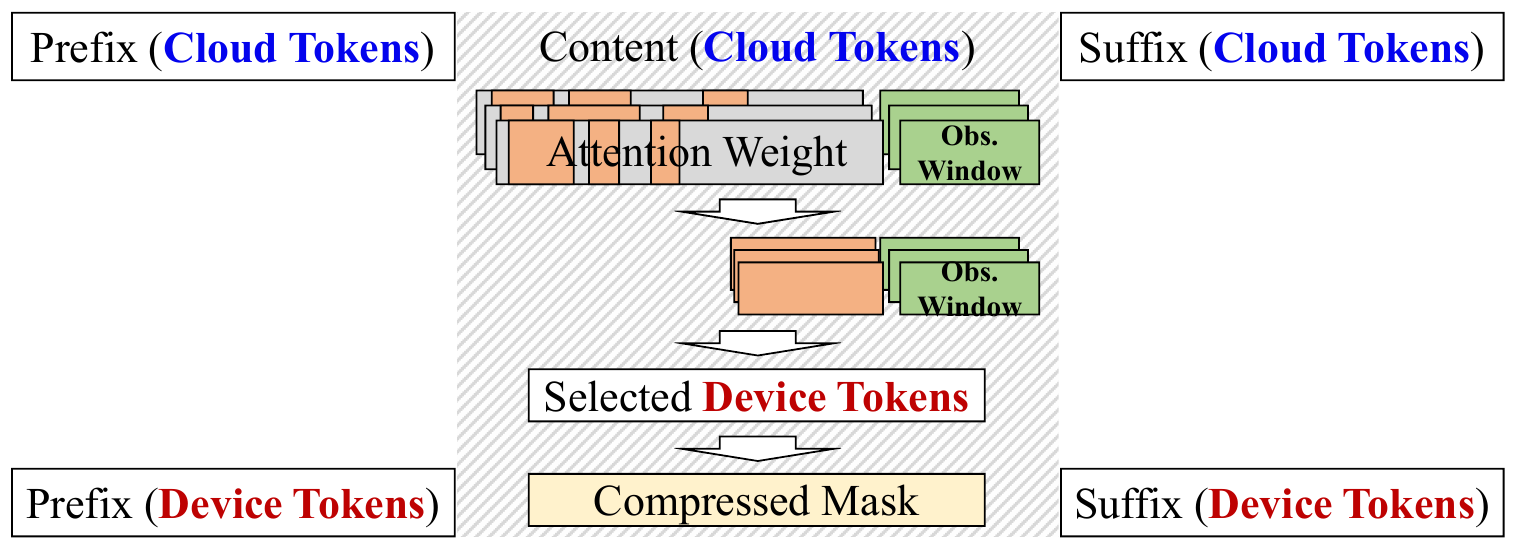}
    \vspace{-17pt}
    \caption{Prompt Refiner in Cloud}
    \label{fig:attention_weight}
    \vspace{-7pt}
\end{figure}

\textbf{Prompt Refiner:}
The refiner is inspired from SnapKV~\cite{LiHYVLYCLC24},
in which each attention head consistently focuses on specific attention features.
Such pattern obtained from an observation window can be used to compress entire KVCache.
However, raw KVCache is unsuitable to be transferred between cloud and devices.
Instead, we convert selected KVCache back to the tokens for device usage.
Unfortunately, directly performing the selection of KVCache may break the semantic coherence.
Therefore, we combine related KVCache in the same sentence, and treat it as a whole.
As a result, the selection of KVCache implies the selection of sentences.
If the model in cloud is the same as that on devices,
the tokens assisted can be directly adopted (or conduct slight correction by the device itself).
Regarding those large models,
their tokenizers may be different from on-device ones.
Thus, the conversion is necessary.
To further compress the data volume, we use KB size mask to indicate selected ones,
as in Fig.~\ref{fig:attention_weight}.

The prompt actually contains three parts: prefix (settings, background), 
content (articles, blogs, etc.) and suffix (questions, format description).
There is no need to refine fixed prefix and suffix.
For content part, related attention weight is obtained along with the implementation of attention
(i.e., $softmax({\text{\textbf{Q}}\text{\textbf{K}}^{\intercal}}/{\sqrt{h}})$, normalized attention score).
Similar to SnapKV, we focus on the last segment of attention weight,
(i.e., part of \textbf{Q} (the observation window) multiplying entire \textbf{K}).
After summing the weight along the query dimension, applying 1D pooling for clustering and sorting,
the indices with top values per head are obtained
(use $r$ to control the ratio of top values selected).
Based on such indices, the positions related to important KVCache are filtered.
Note that all these operations are conducted upon cloud tokens.
We have to convert these selected positions to refined prompt (text) first.
Then, the text is converted to new tokens using device's tokenizer.
The labels in the request indicate the model version.

After refinement, the prompt is about several thousand tokens,
which easily incurs thousands of bytes for transmission (unrealistic for fast feedback).
Note that the refinement essentially filters the tokens (related to selected sentences).
We use a mask tensor to indicate the selection results,
in which ``1'' at position $i$ refers to the confirmation of selection for token $i$.
Instead of transferring such mask tensor,
we further treat it as a bit stream 
and use the compression technique (e.g., gzip) to reduce the data volume.
Via such processing,
the transmission only involves several hundred bytes.
We should mention here that, 
all the operations would be conducted reversely on devices.
Therefore, the compression latency and related resource usage on devices must be controlled.

\textbf{Piggybacking with 1$^{st}$ Token:}
For real-time scenario,
the round-trip communications are desired to be minimized (e.g., exactly once).
To fully utilize the connection,
the compressed mask is transferred via piggybacking, along with the feedback of the 1$^{st}$ token.
For example, the data format is ``token\#mask''.
Since cloud prefill only spends hundreds of milliseconds,
the device receives the 1$^{st}$ token within
\begin{equation}
	\text{TTFT}_{c} = \text{prefill}_{c}(l) + \text{compress}(l, r) + \text{RTT},\label{ttft_c}
\end{equation}
where $l$ is length of original prompt, $RTT$ refers to the round-trip time,
and ``compress'' is the compression time consumed after prefill.
Since the mask size (before compression) is $l$,
``compress'' takes $l$ and refinement ratio $r$ as the input.

Here, we use the attention weight as the guidance.
There are multiple options to be extended 
For example, via the soft prompt~\cite{mu2024learning},
original prompt can be summarized using several tokens,
which can also be transferred along with the 1$^{st}$ token.
Since prompt-level assist is orthogonal to decoding,
its enablement (as plugins) can be easily configured.

\textbf{Early Termination:}
Traditionally, the inference (including both prefill and decoding) 
terminates when end-of-token label (EOT) is generated.
We use $n$ here to represent all $n$ tokens generated.
To control the resource usage in cloud and avoid long occupation,
we use $L$ as the maximum tokens generated ($L$ - 1 decoding tokens), and $L \leq n$.
Two special cases are: 1) $L$ = 1, indicating the cloud only generates the 1$^{st}$ token,
and 2) $L$ = $n$, indicating all decoding tokens are generated.

For intra-cloud requests, the default configuration is $L=*$.
Note that $n$ can only be revealed after the inference 
(various requests may generate different numbers of tokens).
$L=*$ here indicates generating all the tokens.
And for the requests from devices, $L$ is configured per scenario.
During the decoding,
once the number of the tokens generated reaches $L$ - 1,
the inference terminates immediately.
The termination simply refers to release the slot and related KVCache in HBM.
All the tokens, already generated, are transferred to devices 
based on SSE as usual (server push, like a stream).
Thus, no further round-trip communications are needed.

\subsection{On-device Control: Models and Tokens}
\label{sec:design_device}

\textbf{LLM Engine:}
Different from the Ascend stack in cloud, on-device inference is supported by Kirin~\cite{kirinchip}.
As shown before, on-device prefill is quite slower than that in cloud 
(several seconds per thousand tokens).
Instead of conducting the prefill with prompt length $l$ (in tokens),
on-device prefill is postponed until the 1$^{st}$ token is returned 
(also analyzing the mask for shorter prompt).
Note that the length of refined prompt is $rl$.
Meanwhile, upon chunked prefill, the prefix can be treated during TTFT$_{c}$
(other initialization can also be overlapped).
In short, TTFT$_{d}$ is actually the following format:
\begin{equation}
	\text{TTFT}_{d} = \text{TTFT}_{c} + \text{decompress}(l) + \text{prefill}_{d}(rl),\label{ttft_d}
\end{equation}
where ``decompress'' refers to
the time consumed on all reverse operations for recovery from the mask.
Note that user-perceived TTFT is actually TTFT$_{c}$.
TTFT$_{d}$ is used to record on-device TTFT behavior (prefill and extra operations).

Due to limited resources, the device uses some lightweight techniques
(approximation, sparsity, quantification, etc.),
to reduce both computation and memory, which essentially affect the quality.
For example,
on-device attention requires the sliding window implementation, 
in which the window covers at most several thousand tokens.
Although the output shape of attention is $bs * seq * hidden$ $size$,
the inner process involves multiple treatments 
and each treatment covers only part of the input tokens.
Using tensor parallelism or other strategies among NPUs with sufficient HBM,
the cloud is more likely to capture the entire attention behavior.
Enabling the version without lightweight techniques and the precise selection on sentences,
the collaboration achieves a higher quality (compared with on-device inference), 
even using refined prompt and the same model
(shown later in the experiments).

The LoRA is adopted for scenario switching on devices,
in which each LoRA weight (related to the LLM with several billions parameters) 
needs hundreds of megabytes for storage 
and hundreds of milliseconds duration for switching. 

\textit{Remarks:} We should mention here that,
before and during the prompt being transferred to cloud,
the authorization and privacy (the highest priority) are checked and ensured
(authorized by users and APPs, 
removing and substituting sensitive information, supported by private cloud compute, etc.).

\textbf{Speed Controller:}
Although the 1$^{st}$ token is assisted by cloud within TTFT$_{c}$,
waiting for prefill$_{d}$ is still unrealistic (too long, lasting tens of seconds).
Thus, as shown in Eq.~\ref{amortized},
we use $L$ - 1 decoding tokens to amortize on-device prefill.
By reorganizing Eq.~\ref{amortized}, we have ($L$ > 1)
\begin{equation}
    \text{\small TPOT}_{smooth} = \text{\small TPOT}_{d} + \big( \text{prefill}_{d}(rl) - \text{\small TTFT}_{c} \big) / (L-1),\;\label{amortized_v2}
\end{equation}
where long prefill$_{d}$ is amortized to $L$ - 1 decoding tokens.
As shown in previous equation, the value $L$ is needed in advance.
Otherwise \text{TPOT}$_{smooth}$ has to be adjusted dynamically 
during the SSE feedback from cloud.
Thus, the data transferred along with the 1$^{st}$ token,
is further extended to ``token\#mask\#$L$'' 
(or piggybacking within the response body, e.g., in json format).
Note that TTFT$_{c}$ is measured by two timestamps,
between the one for sending the request to cloud and the one for receiving the 1$^{st}$ token.
TPOT$_{d}$ and TTFT$_{d}$ are estimated by the device itself,
in which the bandwidth of accessing SoC memory 
and the input length ($rl$ for prefill and 1 for decoding) are needed.
The most simple method is to multiply a proportional coefficient.
For example, prefill$_{d} \approx k_{d}rl$,
in which $k_{d}$ is the time consumed per unit length on devices
(matching the complexity of sliding window attention).
Further, as shown in previous works~\cite{zhong2024distserve,cheng2025slice} and experiments,
when prompt length is small,
the latency of feed-forward network (FFN) dominates the entire attention 
(given batch size, the complexity of FFN is $\mathcal{O}(l)$).
At the moment of receiving the 1$^{st}$ token
and recovering the refined prompt,
the device uses \text{TPOT}$_{smooth}$ to display follow-up $L$ - 1 tokens from cloud via SSE.

\textbf{Token Corrector:}
If the model deployed in cloud is the same as that on devices, 
the device has the chance to decide the correction policy:
use the cloud one for next generation or use its own for display.
Once the model in cloud is larger (with the same distribution), 
the tokens assisted are regarded as the ``ground truth''.
Here, the ``ground truth'' refers to the correction policy of
using the cloud one for next generation
(as long as the difference occurs at position $i$).

The token corrector and speed controller are orthogonal and can be easily configured.
The controller decides the display speed,
as if they were generated by device itself, using smoothed TPOT.
The token corrector revises the tokens (for display or next token generation),
no matter the display speed.
Since the cloud only helps the device in its prefill phase,
the generation of any decoding token on devices implies
all the tokens from cloud have been ready ($\forall i, i$ < $L$).

Token-level correction is quite simpler than the speculative decoding.
It only compares two tokens at the same position before next generation.
The scope of token corrector is the first $L$ tokens.
Even if the speculative decoding is enabled during the generation,
before or after the verification, the tokens can also be compared, with the ones received from cloud.
Further, two compiled graphs (autoregressive version and verification version) can share the weights on devices.

\textit{Remarks:}
Via token-level assist and prompt level assist,
the model specifications between cloud and devices are essentially decoupled.
The requester on devices has the chance to choose multiple versions of cloud LLMs
while the load balancer in cloud can route the requests 
according to resources and desired rules (ABTest, flow control, etc.).

\subsection{Algorithm on Refiner and Termination}
\label{sec:design_algorithm}

\textbf{Improved TTFT:}
Upon the collaboration, 
cloud TTFT and on-device TTFT are shown in Eq.~\ref{ttft_c} and Eq.~\ref{ttft_d}, respectively,
in which the refinement ratio $r$ is the control variable 
(user-perceived TTFT is TTFT$_{c}$ instead of TTFT$_{d}$).
On-device inference is triggered after the feedback from cloud.
To ensure refined prompt has positive improvement, we have
\begin{equation}
    \text{TTFT}_{d} \leq \text{prefill}_{d}(l),\nonumber
\end{equation}
in which the left depends on the collaboration and the right is traditionally on-device prefill with length $l$.
As mentioned in Section~\ref{sec:design_device}, 
the prefill latency relies on the prompt length.
The simplest estimation is to multiply a proportional coefficient.
Here, we use $k_{c}$ and $k_{d}$ as the coefficients 
(i.e., time consumed per unit length) in cloud and on devices.
As shown in Fig.~\ref{fig:case2_2},
even changing the value of $r$, the extra latency is acceptable and controlled.
Therefore, we use variable $\delta(l)$ as the upper bound of compression, decompression and RTT
(i.e., $\delta(l) \geq $ compress$(l, r)$ + decompression$(l)$ + RTT).
By substituting these variables, we have the following inequality:
\begin{equation}
    \text{TTFT}_{d} \;\leq\; k_{c}l + \delta(l) + k_{d}rl \;\leq\; k_{d}l=\text{prefill}_{d}(l).\nonumber
\end{equation}
By reorganizing it, we have the requirement on $r$ (efficiency):
\begin{equation}
    \xi_{scene} \;\leq\; r \;\leq\; 1 -\; (\;k_{c} + \delta(l)/l\;)\;/\;k_{d},\label{eq_r}
\end{equation}
in which the part $(\;k_{c} + \delta(l)/l\;)/k_{d}$ ``translates'' 
both the cloud prefill time and extra time per unit length 
to the ``time scale'' on devices.
In order to earn all the costs involved from collaboration,
the ratio $r$ should no exceed the right part (less $r$ refers to shorter prompt).
$\xi_{scene}$ refers to the minimum ratio per scenario, to ensure the inference quality.
For example, as mentioned in Fig.~\ref{fig:case2_1},
the quality drops at most several percent ($\xi_{scene}$) under 40\% decrease on prompt length.
 
\textbf{Smoothed TPOT:}
Except for the 1$^{st}$ token,
there are also the requirements on decoding.
In order to control the resource occupation in cloud,
the cloud only helps the devices during its prefill phase.
Therefore, we have (upper bound on $L$)
\begin{equation}
   \text{TTFT}_{c} \;+\; (L-1)*\text{TPOT}_{c} \leq \text{TTFT}_{d}, \nonumber
\end{equation}
in which the left refers to cloud inference (``c'' for cloud).
To ensure the user experience,
the maximum \text{TPOT}$_{smooth}$ should be controlled.
Thus, we have (lower bound on $L$)
\begin{equation}
    \text{TPOT}_{smooth} = \text{\small TPOT}_{d} + \big( \text{prefill}_{d}(rl) - \text{\small TTFT}_{c} \big) / (L-1) \;\leq\; \tau, \nonumber
\end{equation}
in which $\tau$ refers to maximum tolerable TPOT.
As illustrated in Fig.~\ref{fig:case1_2},
with the growth on $L$, \text{TPOT}$_{smooth}$ decreases.
By reorganizing these two inequality, we have
\begin{equation}
    \frac{\text{prefill}_{d}(rl) - \text{TTFT}_{c}}{\tau - \text{TPOT}_{d}} \;\leq\; L - 1 \;\leq\; \frac{\text{TTFT}_{d} - \text{TTFT}_{c}}{\text{TPOT}_{c}}.\label{eq_L}
\end{equation}
Such constraint controls the cloud decoding
(i.e., the number of decoding tokens in cloud should be moderate).
The variable $L$ actually decides the early termination in cloud.
Note that \text{TPOT}$_{d}$ < \text{TPOT}$_{smooth} \leq \tau$,
which amortizes long on-device prefill to the first $L$ - 1 tokens.
And, the rest decoding tokens generated on devices obey \text{TPOT}$_{d}$ (normal TPOT speed).

\setlength{\textfloatsep}{14pt}
\begin{algorithm}[t!]
    \caption{\sysname, Cloud Control}
    \SetAlgoLined
    \let\oldnl\nl 
	\newcommand{\nonl}{\renewcommand{\nl}{\let\nl\oldnl}} 
    \nonl \textcolor{gray}{$\triangleright$ Offline estimation and decision}\\
    Prepare $\tau$, $\xi_{scene}$, $\delta(\cdot)$, $k_{c}$, $k_{d}$, TPOT$_{c}$, TPOT$_{d}$\;
    \For(){<Scene, Device $d$, Prompt Length $l$>}{
        Solve $r$ and $L$ upon $\mathcal{P}$\;
    }
    \nonl ~\\
    \nonl \textcolor{gray}{$\triangleright$ Response per request, token feedback via SSE}\\
    \setcounter{AlgoLine}{0}
    Conduct prefill with length $l$\;
    Obtain $r,L$; Refine and return ``1$^{st}$token\#mask\#$L$''\;
    \For(\textcolor{gray}{$\triangleright$ $L$ controls early termination}){$i \in [1, L-1]$}{
        Conduct decoding, generate token $i$ and return;
    }
    \label{alg:cloud_control}
\end{algorithm}

\textbf{Tradeoff between Efficiency and Quality:}
By combining all previous modeling together, we have the formulation of
\begin{equation}
    [\mathcal{P}]\;\;\;\;Min:\;\; \text{TTFT}_{d} \nonumber,\;\;\;\; s.t.\;\;(\ref{eq_r}),(\ref{eq_L}),
\end{equation}
in which the domain of $r$ is reals ranging from 0 to 1,
and the domain of $L$ is integers larger than 1.
The objective of $\mathcal{P}$ is to minimize on-device TTFT,
since faster TTFT completion leads to earlier start of decoding.
Constraint (\ref{eq_r}) defines the lower bound (quality) and upper bound (efficiency) for $r$.
Constraint (\ref{eq_L}) also defines 
the lower bound (smoothed TPOT) and upper bound (cloud occupation) for $L$.
$\mathcal{P}$ is a mixed integer program.
By using mature libraries, the optimum can be reached,
but the time consumption may fail to match online serving requirement
(i.e., several timeouts during the request lifecycle; at most seconds for TTFT feedback).

\textit{Procedure}s 1 and 2 show the P/D control 
in cloud and on device.
$r$ and $L$ are determined in cloud (optimizer component).
To pursue quick response,
the estimation and decision are made offline.
As in \textit{Procedure} 1,
after preparing all necessary variables and the functions,
the optimizer solves $r$ and $L$ for all possible combinations upon $\mathcal{P}$.
Here, we omit the preparation of functions ``compress'' and ``decompression'',
since both of them can be tested and fitted precisely.
Constraint (\ref{eq_r}) uses the function $\delta(\cdot)$ 
as the upper bound of extra operations.
Both compress and decompress are the functions related to length $l$
(compress is further related to $r$).
$\delta(\cdot)$ also implies the network conditions.
Although precisely estimating RTT is hard,
we categorize it to several common ranges.
For example, via WiFi connections, RTTs are relatively small.
We use the average value to form $\delta(\cdot)$.
As mentioned before, 
we use $k_{*}$ and TPOT$_{*}$ to estimate TTFT$_{*}$ in cloud,
since they can be only revealed after real inference.

\begin{algorithm}[t!]
    \caption{\sysname, Device Control}
    \SetAlgoLined
    \let\oldnl\nl 
	\newcommand{\nonl}{\renewcommand{\nl}{\let\nl\oldnl}} 
    Authorize, desensitize and send request to cloud\;
    Receive 1$^{st}$ token, respond to user; Observe TTFT$_{c}$\;
    \nonl ~\\
    \nonl \textcolor{gray}{$\triangleright$ Parallel branch 1}\\
    Conduct prefill with length $rl$; Observe TTFT$_{d}$\;
    prev $\leftarrow$ 1$^{st}$ cloud token\;
    \While(\textcolor{gray}{\,\,\,$\triangleright$ Decoding with correction}){prev $\neq$ EOT}{
        Decoding, generate token $i$; prev $\leftarrow$ token $i$\;
        prev $\leftarrow$ cloud token $i$, \textbf{if} $i \in \text{\small [1, }L\text{\small-1]}$; Respond prev\;
    }
    \nonl \textcolor{gray}{$\triangleright$ Parallel branch 2}\\
    Use \text{TPOT}$_{smooth}$ to display first $L$-1 decoding tokens\;
\end{algorithm}

The combination considers the triple,
in which the scenario is the most important one.
Note that different scenarios have diverse refinement requirements $\xi_{scene}$
(e.g., some scenarios may not allow the refinement).
Due to the orthogonal feature,
even $r=1$, \sysname still works,
and the improvements gain from token-level assist.
Furthermore, the LLM models used are quite different among scenarios
(different sizes, and would be scaled-out via containers independently).

\begin{table*}[!t]
\fontsize{18}{24}\selectfont
\setlength{\tabcolsep}{13pt}
\centering
\caption{Performance Comparison$^{*}$ on LongBench}\label{exp:longbench}
\vspace{-5pt}
\begin{threeparttable}
\hspace{-10pt}
\scalebox{0.4}{
    \begin{tabular}{l|lcccccccc}
    \specialrule{1pt}{0pt}{2pt}
    &\multirow{3}{*}{~~Methods} & \multicolumn{2}{c}{Baseline} & \multicolumn{4}{c}{Candidates}& \multicolumn{2}{c}{Collaboration} \\
    \cmidrule(lr){3-4}\cmidrule(lr){5-8}\cmidrule(lr){9-10}
    && {Llama~\cite{meta2025llama4}} &{LLMLingua~\cite{jiang2023llmlingua}} &{PyramidInfer~\cite{yang2024pyramidinfer}} &{H2O~\cite{Zhang00CZC0TRBW23}} &{StreamingLLM~\cite{xiao2024efficient}} &{SnapKV~\cite{LiHYVLYCLC24}} &{Refine} &{Refine (Sentences)} \\
        
    \specialrule{1pt}{2pt}{2pt}
    \multirow{5}{*}{\rotatebox[origin=c]{90}{\fontsize{18}{100}\selectfont Single-Doc. QA}}
    &~~~Avg. &25.89 &10.53 &25.40 &20.97 &16.85 &26.15 &26.33 &\textbf{26.64} \\
    \cline{2-10}
    &~~~NrtvQA &18.30 &5.90 &18.96 &15.02 &14.72 &17.58 &18.14 &\textbf{19.60} \\
    &~~~Qasper &20.44 &8.82 &20.11 &17.29 &15.74 &\textbf{21.25} &18.49 &21.16 \\
    &~~~MF-en &34.82 &16.19 &33.72 &29.42 &21.4 &35.32 &35.32 &\textbf{40.95} \\
    &~~~MF-zh &29.99 &11.21 &28.79 &22.16 &15.56 &30.47 &\textbf{33.35} &24.83 \\
        
    \specialrule{1pt}{2pt}{10pt}\specialrule{1pt}{2pt}{2pt}
    \multirow{5}{*}{\rotatebox[origin=c]{90}{\fontsize{18}{100}\selectfont Multi-Doc. QA}}
    &~~~Avg. &22.96 &7.78 &22.39 &19.88 &19.82 &22.43 &\textbf{26.83} &25.83 \\
    \cline{2-10}
    &~~~HotpotQA &29.97 &5.79 &29.73 &28.69 &29.90 &29.96 &\textbf{34.71} &34.22 \\
    &~~~2WikiMQA &27.72 &7.14 &27.49 &25.32 &29.11 &27.75 &\textbf{33.15} &30.99 \\
    &~~~Musique &11.28 &2.87 &11.89 &8.63 &10.64 &10.06 &13.77 &\textbf{14.13} \\
    &~~~DuReader-zh &22.85 &15.33 &20.47 &16.90 &9.63 &21.95 &\textbf{25.70} &23.99 \\
        
    \specialrule{1pt}{2pt}{0pt}
    \end{tabular}
}
\end{threeparttable}
\begin{tablenotes}
    \scriptsize
    \item[] 
    \hspace{-15pt}{\scriptsize *} The models deployed in cloud and on devices are the same (lightweight techniques are adopted on devices). Except for the collaboration, others are test on devices.
\end{tablenotes}
\vspace{-8pt}    
\end{table*}

\textit{Procedure} 2 shows the control on devices.
We should mention here that,
user-perceived TTFT is actually TTFT$_{c}$.
TTFT$_{d}$ is just recorded for calculating \text{TPOT}$_{smooth}$.
When calculating \text{TPOT}$_{smooth}$,
the device still needs estimation.
However, $r$ and $L$ are determined by cloud.
The device only estimates TTFT$_{d}$ itself using $k_{d}$.
Note that length $rl$ can be obtained from the mask.
The procedure is further divided into multiple parallel branches (simultaneously execution).
Branch 2 is responsible for display speed control.
And branch 1 generates decoding tokens after the prefill.
The corrector is implemented via the comparison between two tokens in line 7 (if configured).

\section{Performance Evaluation}
In this section, we conduct empirical experiments to answer the following research questions. 
(RQ1): Whether the cloud-device collaboration effectively improves the LLM inference?
(RQ2): Whether the cloud-device collaboration is efficient and easily scaled out?
And at last, (RQ3): Whether the cloud-device collaboration is friendly to be extended?

\subsection{Prototype and Experimental Settings}

\textbf{Prototype:}
The prompt-level assist is returned along with the first token 
and token-level assist is controlled via early termination.
We implement the cloud functionality as a plugin to existing serving system.
Since the resources are quite different among various devices,
we evaluate the device functionality for either mobile phone or tablet.
Further, it is implemented as the service ability of on-device operating system.

The LLMs in cloud have multiple choices.
The default one is the same as that deployed on devices (with several billion parameters).
Other choices have tens or hundreds of billions of parameters.
Both the Ascend HBM and Kirin SoC memory are tens of gigabytes.
The maximum prompt supported in cloud is hundreds of thousands of tokens
while the one support on devices is several thousand tokens.

Under the user authorization,
the inputs are submitted to a proxy process with users' chatting questions.
The proxy is enabled with multiple inference strategies (configured via UI interaction), 
either on-device inference or requesting for cloud help.
Note that inferring all the decoding tokens is also a strategy of using cloud assistance.
Although both WiFi connections and LTE are supported,
the connections are tested using IP whitelist under WiFi (due to access permission).

\begin{figure}[!t]
    \begin{subfigure}[h]{0.22\textwidth}
        \setlength{\abovecaptionskip}{-1pt}
        \includegraphics[width=1.53in,height=1.265in]{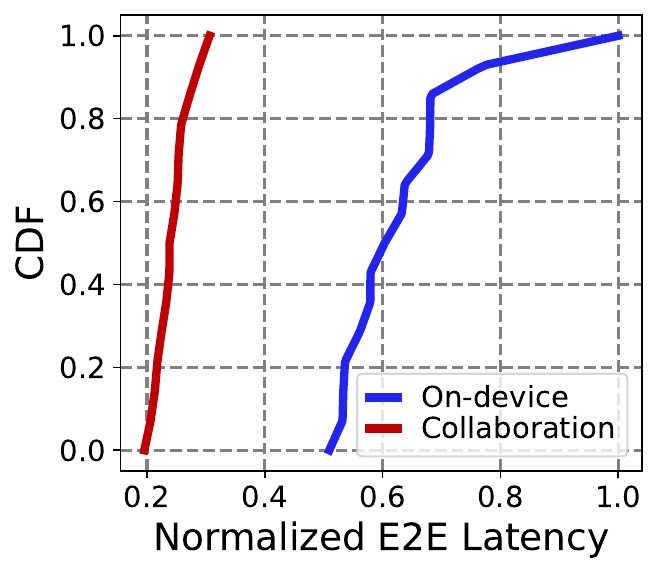}
        \caption{TTFT Results}
        \label{fig:exp1_1}
    \end{subfigure}
    \hfill
    \begin{subfigure}[h]{0.22\textwidth}
        \setlength{\abovecaptionskip}{-1pt}
        \includegraphics[width=1.53in,height=1.265in]{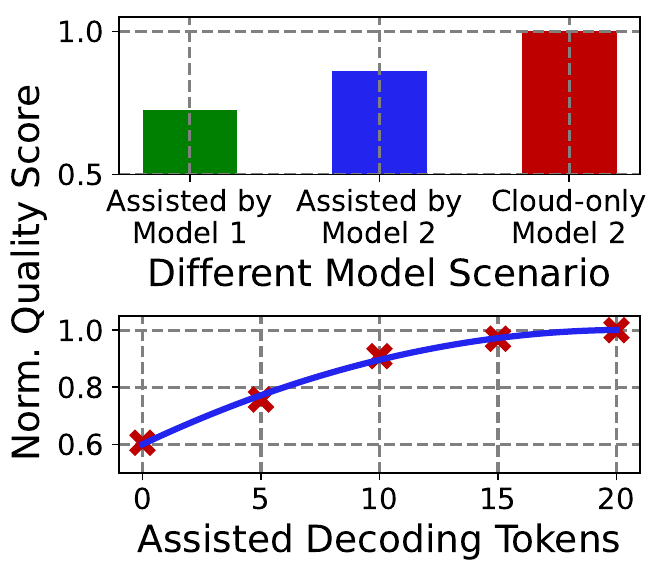}
        \caption{Quality Results}
        \label{fig:exp1_2}
    \end{subfigure}
    \begin{subfigure}[h]{0.22\textwidth}
        \setlength{\abovecaptionskip}{-1pt}
        \includegraphics[width=1.53in,height=1.265in]{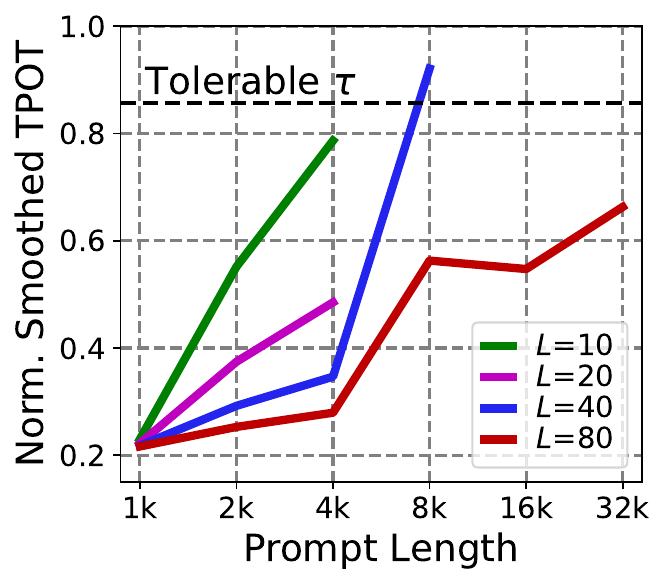}
        \caption{Results of Smoothed TPOT} 
        \label{fig:exp1_3}
    \end{subfigure}
    \hfill
    \begin{subfigure}[h]{0.22\textwidth}
        \setlength{\abovecaptionskip}{-1pt}
        \includegraphics[width=1.53in,height=1.265in]{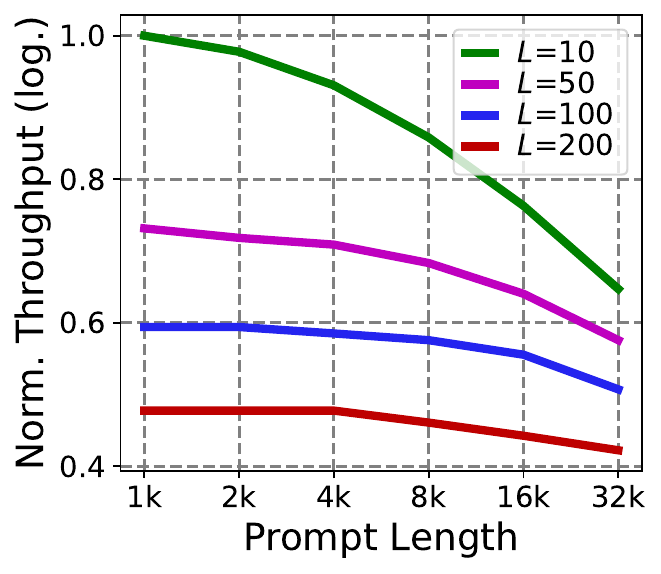}
        \caption{Cloud Throughput Results}
        \label{fig:exp1_4}
    \end{subfigure}
    \vspace{-7pt}
    \caption{Main Metrics during LLM Inference}
    \vspace{-3pt}
\end{figure}

\textbf{Metrics:}
We consider four main metrics: TTFT, TPOT, inference quality and cloud throughput.
Traditional on-device inference performs with long TTFT and a low quality while
cloud inference performs with a low throughput.

\textbf{Datasets:}
The scenarios involve document QA and summarization (major scenarios in LongBench~\cite{bai2024longbench}).
The datasets are organized using diverse contents, 
including technology news, proses, etc.,
in which the maximum length reaches tens of thousands of tokens.
The metric for QA is F1-score.
Meanwhile, other scores like retrieval score, rouge score are involved
(e.g., retrieval one is used when the numbers are in specific formats).
We also use the score measured by humans for bad cases (e.g., semantic inconsistency mentioned before),
which are collected during training and testing.
There are tens of dimensions involved.
For example, mechanical repetition, garbled characters, disorderly citation numbers, etc.
All of the wrong outputs incur deduction on scores.

\subsection{RQ1: Effectiveness of \sysname}
Table~\ref{exp:longbench} shows the performance comparison on LongBench.
Compared with the baseline models (Llama and LLMLingua) and other candidates for prompt or KVCache compression,
the cloud-device collaboration improves the quality.
Specifically, for QA scenarios on either single document or multiple documents,
the collaboration performs the best among majority cases.
And for single document QA, the refinement with sentence level selection performs better.
The prompt length ranges from 4k to 32k, while the default ratio is 0.25.

Fig.~\ref{fig:exp1_1} shows the CDF on the TTFTs for both on-device inference 
(speculative decoding also performs TTFT on devices) and cloud-device collaboration.
Note that the on-device TTFT covers cloud prefill and on-device prefill (with shorter prompt).
Even responding the user using on-device TTFT,
the average latency decreases 60\%.
Moreover, after receiving the first token,
it is directly returned to user (further lower TTFT, user-perceived), 
instead of waiting long on-device inference.
The variance incurred by on-device prefill is large
while that incurred by collaboration is relatively small.
With the growth on the prompt length, cloud TTFT increases slowly,
since the NPUs equipped in the cloud is more powerful.

Fig.~\ref{fig:exp1_2} illustrates the quality 
comparison between the collaboration and cloud-only inference, using different models.
Model 2 is larger than the one deployed on devices.
Cloud-only inference performs the best.
Via collaboration, the quality score drops (about 85\% of cloud-only inference), 
but still better than device-only inference.
Further, with the growth on assited tokens,
the quality score increases (more likely to correct wrong prefix inferred by devices).
In summary, combing Table~\ref{exp:longbench} (same model) and Fig.~\ref{fig:exp1_2} (different models),
the collaboration improves the inference quality (compared with on-device one, and closer to cloud-only inference).

\begin{figure}[!t]
    \begin{subfigure}[h]{0.22\textwidth}
        \setlength{\abovecaptionskip}{-1pt}
        \includegraphics[width=1.53in,height=1.265in]{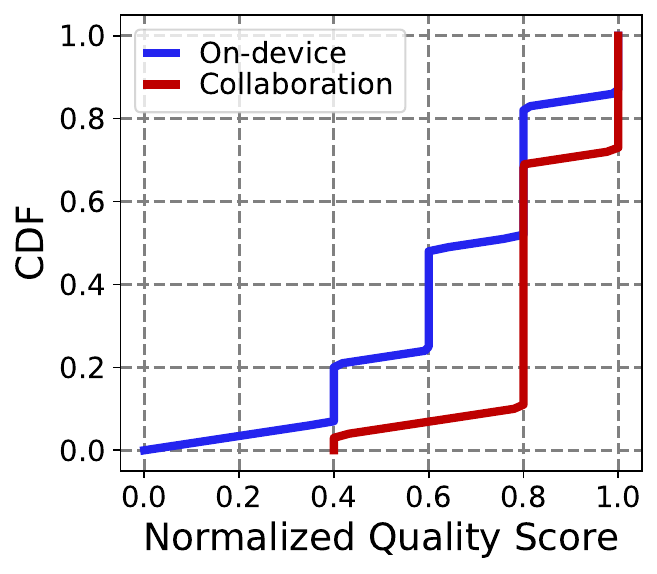}
        \caption{Results Distribution}
        \label{fig:exp0_1}
    \end{subfigure}
    \hfill
    \begin{subfigure}[h]{0.22\textwidth}
        \setlength{\abovecaptionskip}{-1pt}
        \includegraphics[width=1.53in,height=1.265in]{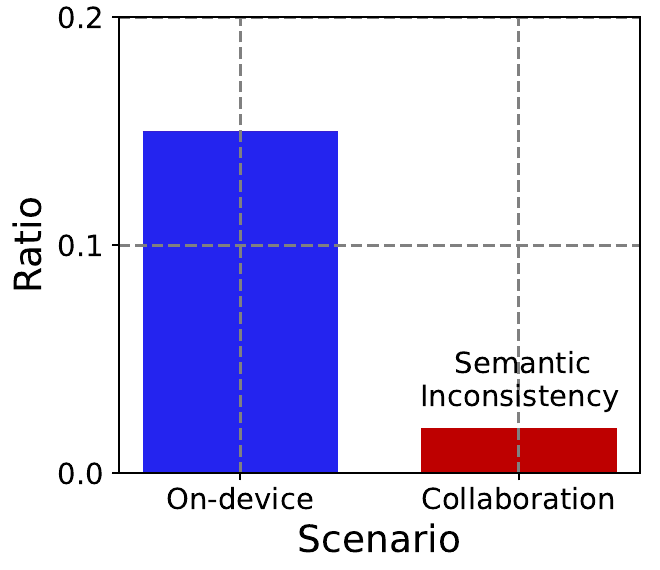}
        \caption{Assisted by a Larger Model}
        \label{fig:exp0_2}
    \end{subfigure}
    \vspace{-7pt}
    \caption{Quality Results on Typical Bad Cases}
    \vspace{-3pt}
\end{figure}

Fig.~\ref{fig:exp1_3} demonstrates smoothed TPOT under various prompt lengths.
The tolerable TPOT (i.e., $\tau$) is about a hundred milliseconds.
When the prompt lengths are less than 4k,
at most 20 tokens are required to amortize long on-device prefill.
Using more tokens assisted leads to smaller TPOT$_{smooth}$ values.
When the prompt length grows to 8k,
40 tokens are moderate (smoothed TPOT is slightly larger than $\tau$, but no harms to the human reading).
As the length further increases,
at most 80 tokens are required (e.g., the optimum one is about 50 tokens for 32k).
The optimum TPOT$_{smooth}$ is calculated according to Constraint (\ref{eq_L}).
In summary, with tens of tokens assisted, the maximum smoothed TPOT is tens of milliseconds.
As shown in the constraint, the optimum is reached 
when $L$ matches the tolerable TPOT (i.e., the lower bound).
Actually, it also indicates the maximum number of token assisted 
(i.e., fully utilize the on-device prefill).
However, with the growth on the tokens assisted, the cloud throughput decreases.

Fig.~\ref{fig:exp1_4} studies the cloud throughput with the changes on $L$ tokens assisted.
Note that the feedback of the first token is a must.
Otherwise, there is no need to send the requests to the cloud.
Compared with 200 tokens generated in cloud,
the cloud-device collaboration improves 1.6x to 15x on cloud throughput
(the prompt length reaches 32k for majority usage),
and the average improvement is 7.6x.
As in Fig.~\ref{fig:exp1_3}, at most tens of tokens are used to smooth TPOT.
Here, the results are evaluated under various $L$ (ranging from 10 to 100),
and the vertical axis is presented in logarithmic form.

As in Fig.~\ref{fig:exp0_1},
even using the refined prompt (prompt-level assist without further decoding tokens), 
the quality score on two bad cases improves 24\%, compared with on-device inference.
Here, the quality scores are measured and divided into several levels.
During on-device inference, the LLMs occasionally generate the outputs with repeated generation and semantic inconsistency.
As a result, several zero scores exist.
The cloud-device collaboration actually improves the minimum level.
Further in Fig.~\ref{fig:exp0_2}, by using a larger model in cloud (even with different output distributions),
the ratio of the bad case (i.e., semantic inconsistency) decreases.

\vspace{-5pt}
\subsection{RQ2: Efficiency of \sysname}
\begin{figure*}[t]
    \begin{subfigure}[t]{0.22\textwidth}
        \setlength{\abovecaptionskip}{2pt}
        \includegraphics[width=1.53in,height=1.265in]{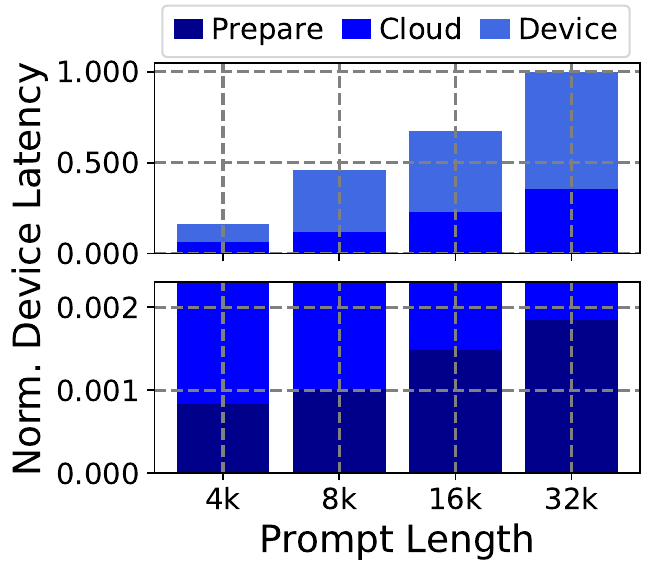}
        \caption{Device Perspective} 
        \label{fig:exp2_1}
    \end{subfigure}
    \hfill
    \begin{subfigure}[t]{0.22\textwidth}
        \setlength{\abovecaptionskip}{2pt}
        \includegraphics[width=1.53in,height=1.265in]{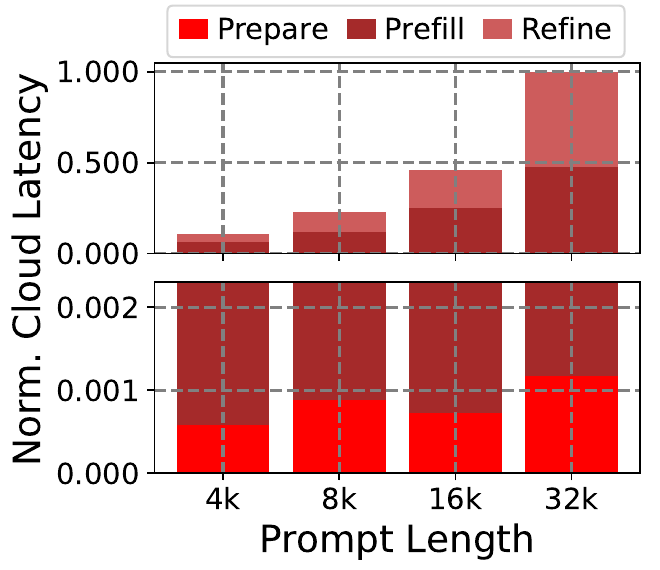}
        \caption{Cloud Perspective}
        \label{fig:exp2_2}
    \end{subfigure}
    \hfill
    \begin{subfigure}[t]{0.22\textwidth}
        \setlength{\abovecaptionskip}{2pt}
        \includegraphics[width=1.53in,height=1.265in]{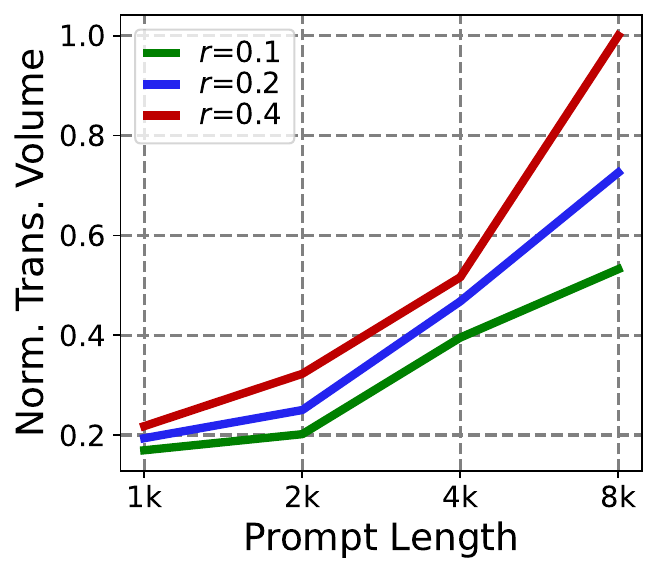}
        \caption{Transmission Volume} 
        \label{fig:exp2_3}
    \end{subfigure}
    \hfill
    \begin{subfigure}[t]{0.22\textwidth}
        \setlength{\abovecaptionskip}{2pt}
        \includegraphics[width=1.53in,height=1.265in]{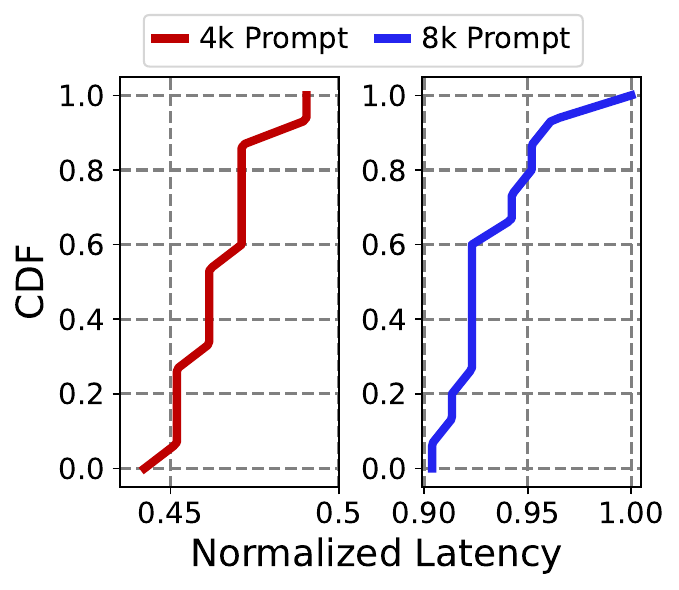}
        \caption{Refinement Latency} 
        \label{fig:exp2_4}
    \end{subfigure}
    \vspace{-7pt}
    \caption{Evaluation on the Efficiency of \sysname}
    \vspace{-1pt}
\end{figure*}

Fig.~\ref{fig:exp2_1} shows the details on the latency, from the perspective of devices.
After tens of milliseconds on preparations, the device requests the cloud for assistance.
The durations between sending the requests and receiving the first token 
range from hundreds of milliseconds to seconds (prompt lengths ranging from 4k to 32k).
After receiving the first token, 
the device further spends seconds on decompression and on-device prefill.
Here, the cloud TTFT contains both cloud operations and RTT.
Note that user-perceived TTFT is lower than on-device TTFT.
And, the completion of on-device TTFT indicates the start of decoding.
Although on-device prefill is larger than cloud prefill,
it has been reduced due to refined prompt.
As mentioned before, on-device TTFT is decreased.

\begin{figure*}[t]
    \begin{subfigure}[t]{0.22\textwidth}
        \setlength{\abovecaptionskip}{2pt}
        \includegraphics[width=1.53in,height=1.265in]{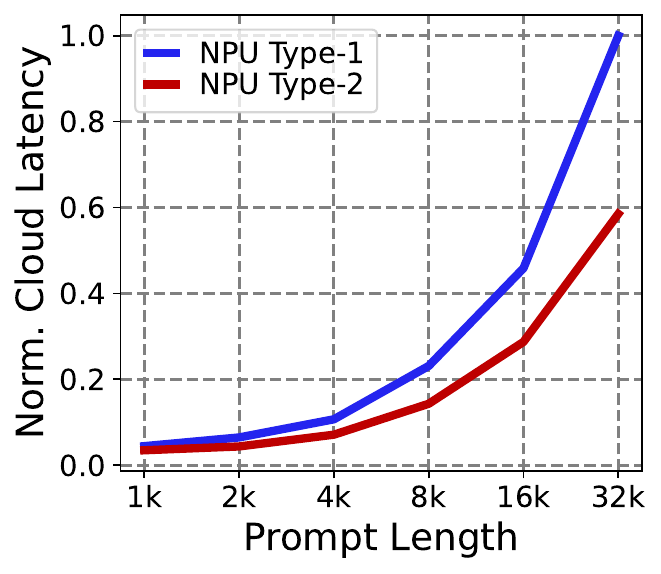}
        \caption{Using Various NPUs} 
        \label{fig:exp3_1}
    \end{subfigure}
    \hfill
    \begin{subfigure}[t]{0.22\textwidth}
        \setlength{\abovecaptionskip}{2pt}
        \includegraphics[width=1.53in,height=1.265in]{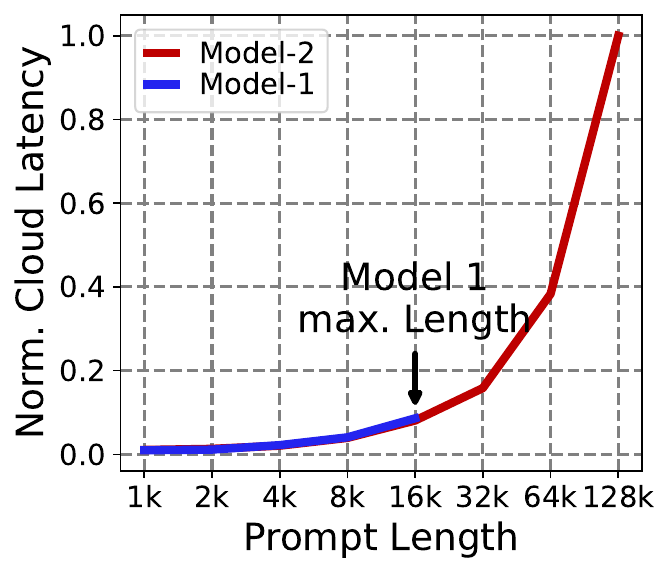}
        \caption{Using Various Models}
        \label{fig:exp3_2}
    \end{subfigure}
    \hfill
    \begin{subfigure}[t]{0.22\textwidth}
        \setlength{\abovecaptionskip}{2pt}
        \includegraphics[width=1.53in,height=1.265in]{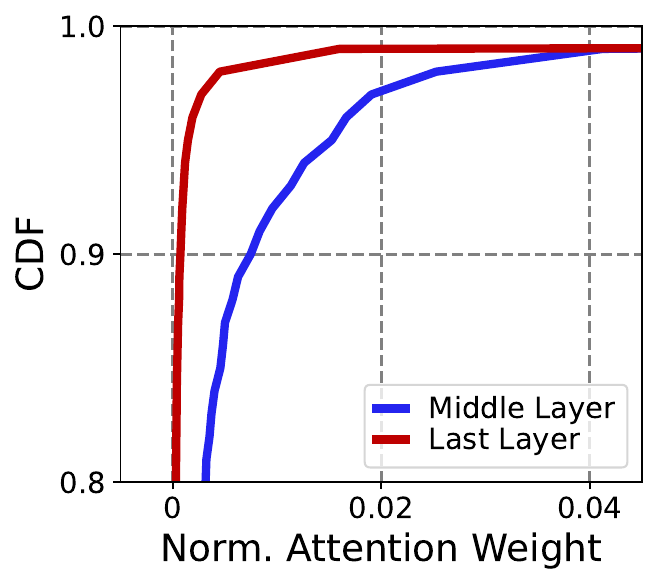}
        \caption{Attention Distributions} 
        \label{fig:exp3_3}
    \end{subfigure}
    \hfill
    \begin{subfigure}[t]{0.22\textwidth}
        \setlength{\abovecaptionskip}{2pt}
        \includegraphics[width=1.53in,height=1.265in]{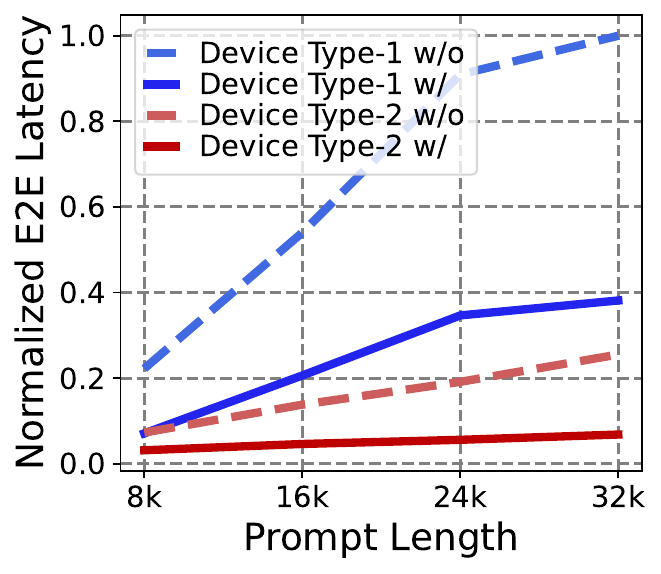}
        \caption{Using Various Devices} 
        \label{fig:exp3_4}
    \end{subfigure}
    \vspace{-7pt}
    \caption{Evaluation on the Compatibility of \sysname}
    \vspace{-6pt}
\end{figure*}

Fig.~\ref{fig:exp2_2} illustrates the details on the latency, from the perspective of cloud.
The cloud also spends several milliseconds on preparations and forwarding.
Note that all the instances are deployed using containers.
Therefore, the cost is a must, 
including forwarding among SLB (service level balancer) and gateway, batching, logging metrics, etc.
About half of the time is used on cloud prefill inference.
And the rest is spent on refinement, including the operations on mask and compression.
The cloud prefill can be optimized using various parallelism strategies.
For example, the prefill here is conducted within one node.
Further sequence parallelism is enabled to achieve lower TTFT for long prompt, 
with more nodes involved. 

Fig.~\ref{fig:exp2_3} demonstrates the data volume transferred from the cloud to devices.
The refinement ratio actually controls the prompt length, according to its definition.
Given one prompt, with the decrease on refinement ratio, 
the data volume also decreases (i.e., the data after refinement, including selection and compression).
With the growth on the prompt length, the data volume transferred increases.
Since refined prompt is appended, piggybacking with the feedback of the first token,
the data volume here refers to the string length in the response body (formatted in json).
Each follow-up decoding token only incurs tens of strings for transmission
(before the early termination).
In summary, the maximum data volume is hundreds KBs (8k prompt) along with the first token.

Fig.~\ref{fig:exp2_4} studies the CDF on refinement latency for various prompt lengths.
The latency is almost proportional to related prompt length.
Given fixed prompt length,
the variance of the latency is small (also shown in Fig.~\ref{fig:case2_2}).
However, with the growth on prompt length,
the variance increases accordingly.
The maximum latency on refinement for 8k prompt is several hundred milliseconds
(under heavy load, the proportion is small).
The majority refinement latency for 8k prompt is about one or two hundred milliseconds (acceptable).

\subsection{RQ3: Compatibility of \sysname}

Fig.~\ref{fig:exp3_1} shows the inference latency under various NPUs (in cloud).
Note that the data transferred between the cloud and devices is strings (refined prompt and tokens).
Therefore, the internal details of the cloud on implementation and NPU types are shielded.
Actually, multiple types of NPU are deployed in cloud.
Then, the cloud is expected to route the requests to 
different services with separate model sizes and NPU types for various SLOs.
Even within the same series,
the computing capacity, HBM, etc., are different.
For example, in terms of the TFLOPs,
the maximum one is several times greater than the minimum one.
The cloud considers the combinations of NPUs, parallelism strategies and models for various scenarios
(optimized combination is not within the scope of this paper).
Currently, the disaggregation of prefill and decoding, or expert parallelism,
are conducted using the same types of NPUs.
Further collaboration among different NPUs is exploring.

Fig.~\ref{fig:exp3_2} illustrates the compatibility for different models.
The serving system is easily extended to multiple model sizes and series.
Some of them pursue high quality while the others pursue the inference speed.
For example, the maximum prompt length supported by model 1 is small (only 16k) in the figure.
Its inference speed is also slower than that of model two.
However, the quality of it is higher than the others.
By using only one node, maximum prompt length supported by model two is 128k.
As mentioned before, equipped with sequence parallelism, tensor parallelism among multiple nodes,
the maximum prompt length could be extended.

Fig.~\ref{fig:exp3_3} demonstrates the attention weights (selected ones)
of the middle layer and the last layer (models with tens of layers).
Similar to SnapKV, some of the attention layers are involved.
However, different layers show quite different behaviors on the distribution.
For example, the ones selected in the middle layer are larger than that in the last layer.
Therefore, there are multiple strategies on the layer selection, as well as related window size.
Further, the values of attention weights per layer are relatively small,
but the variance is large.
Here, the majority values are only several percent of the maximum one.
By considering the difference among layers and the consistence per layer,
multiple approaches can be further adopted.
These approaches are orthogonal to the prefill.

Fig.~\ref{fig:exp3_4} studies the results using various devices.
The red lines are performed using tablets while the blue lines are performed using mobile phones.
Via enabling the collaboration, the end-to-end (E2E) latency dramatically decreases.
Since the computing capacity on tablet is stronger than that on mobile phone.
The latency using tablet is lower than that on mobile phone, no matter the collaboration is enabled or not.
By using the collaboration, the latency gap among various devices decreases.
The improvement on collaboration is large,
when the prefill is conducted on weak devices.

\section{Related Works and Extensions}

\textbf{Cluster-scale and On-device LLMs:}
Numerous related references have emerged within a short span.
They focus on either cluster-scale serving system or on-device LLM inference.

Serving LLMs in a disaggregated paradigm is a new trend.
Splitwise~\cite{patel2023splitwise} and DistServe~\cite{zhong2024distserve}
proposed to place prefill and decoding phases on different devices
to prevent related interference (compute-bound prefill and memory-bound decoding).
Mooncake~\cite{qin2025fast} and P/D-Serve~\cite{jin2024pdserve} 
demonstrated the disaggregated LLMs over thousands of GPUs or NPUs, with MLOps and related service management.
These works~\cite{gao2024attentionstore,qin2025fast,zuo2025serving} used distributed KVCache pools for achieving various SLOs.
P/D-Serve and the practice on CloudMatrix384 used directly device-to-device network to transfer KVCache.
TetriInfer~\cite{cunchen2024tetriinfer} scheduled the two phases to shorten the execution timeline.
There are also some works focusing on 
prefix caching~\cite{ye2024chunkattention,0002HQZYCZH025}, 
chunked prefill~\cite{agrawal2023sarathi,ZengGLYSHWZ00Q24,qiao2025swiftkv} and 
various parallelism strategies~\cite{brakel2024modelparallelism,wang2025flexsp,lin2025apex,deepseek2024v3},
to accelerate LLM inference.

On-device LLM focuses on accelerating the inference using fast attention, quantification, pruning, etc.
Some works studied the sparsity of attention~\cite{child19swa,gao2025seerattention,SongCYDGCS24,yang2025lserve}.
Longformer~\cite{beltagy2020longformer}, BigBird~\cite{zaheer2021bigbird} and Attention Sink~\cite{xiao2024efficient}
used sliding window to reduce the computation of attention (tokens involved).
Other works explored to use the approximation operations~\cite{KatharopoulosV020,altabaa2024approximation},
caching and well-designed pipelines to facilitate better performance 
on target hardware~\cite{Dao24,ShahBZTRD24,xue2024powerinfer2fastlargelanguage,tan2024pipellm,WangGZYG25}.
Some works like Any-Precision LLM~\cite{park2024anyprecision,kim2025guidedquant,chen2025progressive,liu2025flexquant} 
proposed lightweight methods for precision quantization.
And, some works like~\cite{FrantarA23,ZhengCQSSC25,yi2025symmetricpruning,ye2025oneforallpruning} 
investigated the pruning techniques for LLMs.
Note that the speculative decoding is also 
enabled and accelerated~\cite{xu2023llmcad,cai2024medusa,SvirschevskiMCC24,XuYZJZWXL25,hu2025speculative}.

\textbf{Cloud-device Collaboration:}
For LLMs, numerous works focus on the model partitioning for collaborative inference.

Edge AI~\cite{LiZZC20}, DeepSlicing~\cite{ZhangZQWJL21}, EdgeShard ~\cite{zhang2024edgeshard}, 
etc.~\cite{LaskaridisVALL20,AlmeidaLVLL22,ChenLZPWHZ24},
partitioned the DNN models 
among devices and the sites (edges, base station, cloud, etc.).
Raw intermediate data during inference has to be transferred.
Hao $et\;al.$~\cite{HaoJJ0C24} proposed edge-cloud collaboration for speculative decoding (token-level transmission).
However, all the candidates have to be verified in time.
Venkatesha $et\;al.$~\cite{venkatesha2025fastcost} proposed a fast and cost-effective 
speculative edge-cloud decoding framework, in which the pipeline was well orchestrated. 

There are also some works for LLM and SLM collaboration~\cite{HaoJJ0C24,bergner2024thinkbig,kim2025guiding}, 
mainly in cluster.
Although the GKT~\cite{yao2024gkt} considered related cloud-edge mechanism, 
the input of SLM couldn't be long (otherwise, long on-device prefill involved),
which was unsuitable for the content with much details.

Other works study the KVCache compression, selection as well as related prompt summary.
SnapKV~\cite{LiHYVLYCLC24} automatically compressed KVCache by selecting clustered important KV positions per head.
H2O~\cite{Zhang00CZC0TRBW23} dropped part of KVCache during the generation upon a scoring function.
FastGen~\cite{Ge0LZ0024} studied various KVCache compression strategies.
Gisting~\cite{mu2024learning} trained the model to compress prompts into smaller sets of gist tokens.
These works effectively compress the KVCache and related prompt.
However, the data volume to be transferred should be controlled for real-time communication.

\textbf{Cloud Thinks and Device Acts:}
The cloud only helps a portion of the content for each device,
as long as the information is produced during the prefill.
Along with the 1$^{st}$ token,
other information like planning results, templates, agent controls, etc., 
can also be used to guide the device execution.
\section{Conclusion}
This paper proposes a realistic cloud-device collaboration mechanism for LLM,
in which the cloud helps a portion of the content for each device, only in its prefill phase.
With the token-level assist, 
the device uses them to amortize long TTFT, leading to a smoothed TPOT.
And after catching up the progress, the device generates tokens itself.
Furthermore, during the cloud prefill, the cloud refines the prompt,
and the information is transferred via the piggybacking with the 1$^{st}$ token.
We also propose an algorithm to decide the best settings.
We implement such scheme \sysname in our prototype 
and confirm the superiority over other alternatives.

\bibliography{main}
\bibliographystyle{unsrt}
\balance

\end{document}